\documentclass[aoas,nameyear,dvips]{arximspdf}
\usepackage{dcolumn}
\usepackage{subenv}
\usepackage{multirow,stfloats}
\usepackage{graphicx}

\doi{10.1214/10-AOAS403}
\volume{5}
\issue{2A}
\pubyear{2011}
\firstpage{843}
\lastpage{872}

\makeatletter
\newcolumntype{d}[1]{D{.}{.}{#1}}
\newcommand{\Cov}{\operatorname{Cov}}
\renewcommand{\epsilon}{\varepsilon}
\fnbelowfloat  % prie global komandu
\makeatother

\begin{document}
\begin{frontmatter}

\title{A mixed effects model for longitudinal relational and network
data, with applications
to international trade and conflict}
\runtitle{Longitudinal relational data}

\begin{aug}
\author[A]{\fnms{Anton H.} \snm{Westveld}\corref{}\ead[label=e1]{anton.westveld@unlv.edu}}
\and
\author[B]{\fnms{Peter D.} \snm{Hoff}\thanksref{t3}\ead[label=e2]{pdhoff@uw.edu}}
\runauthor{A. H. Westveld and P. D. Hoff}
\affiliation{University of Nevada, Las Vegas and University of
Washington, Seattle}
\address[A]{Department of Mathematical Sciences\\
University of Nevada, Las Vegas\\
Las Vegas, Nevada\\
USA\\
\printead{e1}} %adresu isvedimo komanda gale!
\address[B]{Department of Statistics\\
University of Washington, Seattle\\
Seattle, Washington\\
USA\\
\printead{e2}}
\end{aug}

\thankstext{t3}{Supported in part by NSF Grant SES-0631531.}

% HISTORY:
\received{\smonth{7} \syear{2009}}
\revised{\smonth{8} \syear{2010}}

% ABSTRACT
%
\begin{abstract}
The focus of this paper is an approach to the modeling of longitudinal
social network or relational data.
Such data arise from measurements on pairs of objects or actors
made at regular temporal intervals, resulting in a social network for
each point in time.
In this article we represent the network and temporal dependencies
with a~random effects model, resulting in a stochastic process defined
by a set of stationary covariance matrices.
Our approach builds upon the social relations models of Warner, Kenny
and Stoto
[\textit{Journal of Personality and Social Psychology}
\textbf{37} (1979) 1742--1757] and
Gill and Swartz [\textit{Canad. J. Statist.} \textbf{29} (2001) 321--331]
and allows for an
intra- and inter-temporal representation of network structures.
We apply the methodology to two longitudinal data sets: international trade
(continuous response) and militarized interstate disputes (binary response).
\end{abstract}

% KEYWORDS
%

\end{frontmatter}
\setcounter{footnote}{1}

%s1 ###
\section{Longitudinal network (relational) data}
\label{sec1}

Radcliffe-Brown
(\citeyear{radcliffebrown1940})
stated that an understanding of the
``complex network of social relations'' can be gained by measuring the
relations or interactions within a set of actors. Since pairwise
relations are the most elemental type of relationship, relational data
which consist of measurements made on pairs of actors are ubiquitous.
Our focus in this article is on relational data from the field of
political science, including (1) trade between nations, and (2)
militarized disputes between nations. For such data, we let $y_{i,j}$
denote the value of the measurement on the potentially ordered pair of
actors ($i,j$). In this paper we refer to \textit{social network data}
or \textit{relational data} as the set of measurements of relations on
dyads for a group of actors under study. These measurements could be
binary, ordinal or continuous, as such, the methodology applies to a
broad range of applications beyond those discussed in this paper.

In the case of international trade, $y_{i,j}$ is the directed level of
trade from nation $i$ to nation $j$. Since the relation is directed,
$y_{i,j}$ is not necessarily equal to $y_{j,i}$. Typically, social
network data, directed or undirected, are represented by a \textit
{socio-matrix} [\citet{Was94}], with the $i$th row representing data for
which actor $i$ is the sender, and column $j$ representing data for
which $j$ is the receiver. Since the data are based on pairs of actors,
the diagonal representing the relationships of actors with themselves
is generally absent from the socio-matrix.

Many researchers have worked on models for this data structure. The
seminal work on relational data of this form was done by \citet
{WarKenSto79}, where a method of moments estimation procedure was
developed based upon an ANOVA style decomposition. Models of this form
have come to be known as \textit{social relations} models or models for
\textit{round robin} data. \citeauthor{Won82}'s (\citeyear{Won82}) work derived maximum
likelihood estimators for these types of models, and \citet{GilSwa01}
studied method of moments, maximum likelihood and Bayesian estimation
procedures for the same problem. More broadly, \citet{HenLi02} and
\citet{LiLok02} developed a general unified theory for dyadic data
which derives the social relations model and other similar models from
principles of group symmetry and exchangeability.

In a series of papers [\citet{Hof02};
\citeauthor{Hof03} (\citeyear{Hof03,Hofbi,HofSVD})], the social
relations model was expanded in several directions: (1)~A~la\-tent social
space was introduced to capture patterns of transitivity, balance and
clusterability that are often exhibited in dyadic data [\citet{Was94}];
(2) A generalized linear model was developed to allow for a variety of
data types (binary, ordinal and continuous); (3)~A~Baye\-sian estimation
procedure was thoroughly outlined for (1) and (2) to estimate the model
parameters.

However, all models mentioned thus far are for static relational data.
Often, scientific questions are concerned with the evolution of
networks over time. For example, in the field of international
relations, questions related to the evolution of international trade or
interstate conflicts are of great interest [\citet
{HofWar03}; \citet{WarHof07}; \citet{WarSivCao06}]. In the field of biology, an
understanding of the evolution of interactions of biological entities
under various experimental stimuli could provide important insights
[\citet{BarOlt04}].
With such applications in mind, this paper expands the social relations
model to account for dependence over time.

This article proposes a model that accounts for temporal dependence
among all pairwise measurements of a set of actors, thus, it falls into
the realm of longitudinal data analysis methodology. To date, there has
been little work on models which account for both network and temporal
dependencies. A notable exception is the work by Thomas Snijders and
coauthors [Huis\-man and Snijders\vadjust{\eject} \citeyear{HusSni03}; \citet{SniBun09}; \citet{SniKosSch10}] which developed
an~\mbox{actor-orien-}\break ted model for network evolution that incorporated
individual-level attributes. This approach is based on an economic
model of rational choice, whereby individuals make unilateral changes
to their networks and behaviors in order to maximize personal utility
functions. Parameter estimates describe individual's utilities for
various network configurations. Parameter estimation methods for such a
model have been developed into a freely-available software package (\url
{http://stat.gamma.rug.nl/siena.html}), which has been applied to a
number of data sets.

While this work has been groundbreaking, the applicability of an
actor-oriented model may be limited to certain types of networks. As
described by the primary developers of this approach [\citet{SniSte07}],
such a model may not be appropriate in situations for which network and
behavioral data might depend on unobserved latent variables.
Additionally, the interpretation of the parameters in an actor-based
choice model may be problematic if the data do not actually represent
choices of the actors, but rather outcomes
determined by other actors, which may be constrained by circumstances
beyond an individual's control. In the context of trade, for example,
exports from one country to
another may be determined by forces of supply and demand beyond just
the pair. In the context of international conflict, countries are often
unwilling participants in militarized disputes.

Recently, \citet{HanFuXin10} considered a temporal extension of the
\textit{exponential random graph} modeling (ERGM) framework [\citet
{FraStr86}; \citet{HunHan06}; \citet{HanRafTan07}]. Their work is similar to that of
Thomas Snijders and coauthors in that it parameterizes various network
configurations, however, they do not take an agent-based approach to
the construction of the model.

Another approach to modeling dynamic network data is
discussed in \citet{XinFuSon10}.
Building on ideas of \citet{erosheva2004} and \citet{airoldi2005},
these authors model each actor as
having partial memberships to several groups. Relationships between
individuals are determined by the groups of which they are members.
Such models often result in a concise description of the data,
as the large
number of relationships between actors
are summarized by the relationships between a~small number of groups
to which the actors belong.

In contrast to an actor-oriented utility model, ERGM, or a
group-member\-ship model, the approach we propose is more statistical, in
that the main parameters in our model represent expectations and
covariances of relational measurements leading to inference about
network characteristics. Our reason for taking such an approach is that
in the empirical study of international
relations, focus is primarily on mean or regression effects and assessments
of their statistical significance. As discussed in \citet{WarHof07}, common
practice is to merge data on all pairs of countries across several years
and base inference on ordinary least squares estimates, treating
all observations as independent. By ignoring network and temporal
dependence, such an approach can potentially
dramatically overestimate the significance of results and precision of
estimates. One of the objectives of our methodology is to provide
mean and regression estimates, by properly accounting for
statistical dependencies in the data.
Additionally, our modeling framework is very flexible and extendable:
Using a generalized linear model framework, it can accommodate
continuous and ordinal relational data. This could include data on
intensity or duration of relationships, or the number of contacts
between two individuals.

The next section will outline the set of possible second-order
dependencies \mbox{inherent} in this data structure. Section~\ref{sec3} represents the
dependencies with a~mixed-effects model, and Section~\ref{sec4} outlines a
Bayesian approach to parameter estimation. Sections~\ref{sec5} and~\ref{sec6} provide
in-depth data analysis examples involving international trade and
militarized disputes, including comparisons to simpler modeling
approaches. A discussion follows in Section~\ref{sec7}.

%%%%%%%%%%%%%%%%%%%%%%%%%%%%%%%%%%%%%%%%%
%s2 ###
\section{Dependence structure for LSR data}\label{sec2}

\begin{figure}
\centering
\begin{tabular}{@{}ccc@{}}

\includegraphics{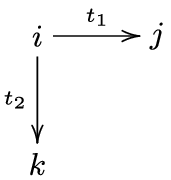}
&\includegraphics{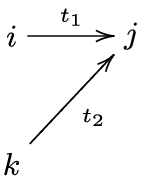}&\includegraphics{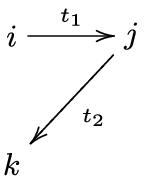}\\
\footnotesize{(a) same sender}&\footnotesize{(b) same receiver}&\footnotesize{(c) common participant}\\[6pt]

\includegraphics{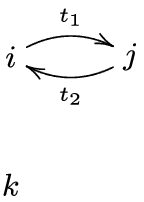}
&\includegraphics{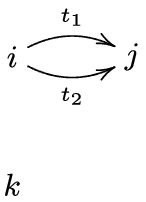}\\
\footnotesize{(d) reciprocity}&\footnotesize{(e) observational dependence}
\end{tabular}
\caption{Second order dependencies for longitudinal directed social
network data.}\label{DepStr}
\end{figure}

Figure \ref{DepStr} summarizes the set of pairwise (second order)
potential dependencies for directed longitudinal social network data
that we will consider in this article. These are the dependencies
possible assuming a dependence
structure
in which two relational measurements are dependent if and only
if they share a common actor.
In the figure, three nonidentical actors $i,j,k$ and two time points
$t_1, t_2$ are used to illustrate the dependencies. The arrow $
i \stackrel{t}{\longrightarrow} j $ represents the random variable $y_{i,j,t}$ for a
particular relationship from actor $i$ to actor $j$ at time $t$. If we
are to study patterns of international trade, $i \stackrel{t}{\longrightarrow} j $ might represent the monetary value of the exports from nation $i$
to nation $j$ during year $t$. Based on the figure, two directed
relations are potentially dependent only if they share a common actor,
regardless of the relation's time index. In other words, the random
variables $y_{a,b,t_1}$ and $y_{c,d,t_2}$ are independent for all $t$
if $\{a,b\} \cap\{c,d\} = \varnothing$.

%f1

To provide a description of the five dependencies depicted in Figure
\ref{DepStr}, let us first consider a fixed time point $t_1=t_2=t$.
Under this condition, (a)~represents the potential dependence among
measurements having a
common sender (i.e.,  the ``row effects''). As an example of such dependence,
consider the exports from the United States and those
from Morocco in a given year.
Due to the overall difference in trade activity of these two nations,
we might expect the exports from Morocco to other countries to be more
similar in magnitude to each other than to the exports from the United
States to other countries. Similarly, (b) represents potential
dependence among measurements having a common receiver (i.e.,  the\vadjust{\goodbreak}
``column effects''). Considering the context of international trade
again, some countries consume more goods than other countries, which
could lead to within-column correlation of trade values. Next, (c)
represents dependence between the relations sent and received by the
same actor. For example, countries that import at a higher than average
rate may also export at a higher rate. Next, (d) represents the idea of
reciprocity or dependence between the directed relations of a pair of
actors, such as between a pair of international trading partners or
disputes between a pair of nations.

For $t_1 \not= t_2$ we additionally consider temporal dependence: The
figure in~(e) indicates the dependence among a pair of actors across
time.\footnote{In the case of nondirected network data, the minimal set
of dependencies are obtained by replacing the directed edges in Figure
\ref{DepStr} with nondirected edges. In this case, cases (a), (b) and
(c) essentially represent
the same dependencies, as do (d) and (e). This leaves only two minimal
dependencies.}

%%%%%%%%%%%%%%%%%%%%%%%%%%%%%%%%%%%%%%%%%%%
%s3 ###
\section{Mixed effects model with Markov temporal dependence}\label{sec3}
We base our longitudinal network model on
multivariate normal distributions,
with nonzero covariances corresponding to the
dependencies represented by Figure \ref{DepStr}. For example,
we allow $\Cov(y_{a,b,t_1}, y_{c,d,t_2})\not= 0$ if
$\{a,b\} \cap\{c,d\} \not= \varnothing$.
Otherwise, this covariance is zero.
Based on this, the complete set of nonzero covariances are in Table \ref
{GenCovStr}.
Such a covariance structure can be
obtained via a mixed effects model, defined by equations (\ref
{ModEqn})--(\ref{ModStaCovEqn}) that follow:
%
%e1 ###
\begin{eqnarray}
\label{ModEqn}
y_{i,j,t} &=& x_{i,j,t}'\beta_t + \epsilon_{i,j,t}, \nonumber
\\[-8pt]
\\[-8pt]
\epsilon_{i,j,t} &=& s_{i,t} + r_{j,t} + g_{i,j,t}.
\nonumber
\end{eqnarray}
In this model, $x_{i,j,t}'\beta_t$ is a fixed effect expressing the
mean for $y_{i,j,t}$, while the error term $\epsilon_{i,j,t}$ is
decomposed into a set of mean-zero Gaussian random effects. This linear
decomposition consists of a sending effect $s_{i,t}$, a receiving
effect $r_{j,t}$ and a residual error term $g_{i,j,t}$. For a fixed
$t$, the network dependencies can be characterized by specifying
covariance structures for the random effects in (\ref{ModEqn}).

%
%t1 ###
\begin{table}%[!htb]
\caption{Covariances in the longitudinal social relations model}
\label{GenCovStr}
\begin{tabular}{@{}llcc@{}}
\hline
& &  $\bolds{t_1=t_2}$ & $\bolds{t_1<t_2}$ \\
\hline
(a)  &  $\Cov( i \stackrel{t_1}{\longrightarrow} j, i
\stackrel{t_2}{\longrightarrow} k) =  \Cov(y_{i,j,t_1},y_{i,k,t_2}) =$
&$\xi^2_{1,
[t_1]}$ & $\xi_{1,[t_1,t_2]}$ \\
[3pt]
(b)&  $\Cov( i \stackrel{t_1}{\longrightarrow} j, k
\stackrel{t_2}{\longrightarrow} j) =   \Cov(y_{i,j,t_1},y_{k,j,t_2}) =$ & $\xi^2_{2,
[t_1]}$ & $\xi_{2,[t_1,t_2]}$\\
[3pt]
\multirow{1}{10pt}[-6pt]{(c)}&
$\Cov( i \stackrel{t_1}{\longrightarrow} j , j \stackrel{t_2}{\longrightarrow}
k )  =\Cov(y_{i,j,t_1},y_{j,k,t_2}) = $&$\xi_{3,[t_1]}$&$\xi_{3,[t_1,t_2]}$\\
&$\Cov( i \stackrel{t_2}{\longrightarrow} j , j \stackrel{t_1}{\longrightarrow}
k)  =\Cov(y_{i,j,t_2},y_{j,k,t_1}) =$&$\xi_{4,[t_1]}$&$\xi_{4,[t_1,t_2]}$\\
[3pt]
(d)& $\Cov( i \stackrel{t_1}{\longrightarrow} j, j
\stackrel{t_2}{\longrightarrow} i) =   \Cov(y_{i,j,t_1},y_{j,i,t_2}) =$ & $\xi
_{5,[t_1]}$ & $\xi_{5,[t_1,t_2]}$ \\ [3pt]
(e)&   $\Cov( i \stackrel{t_1}{\longrightarrow} j, i
\stackrel{t_2}{\longrightarrow} j) =   \Cov(y_{i,j,t_1},y_{i,j,t_2}) =$ & $\xi^2_{6,
[t_1]}$ & $\xi_{6,[t_1,t_2]}$
\\
\hline
\end{tabular}
\end{table}

While Figure \ref{DepStr} describes the structure of the network dependencies
(pairwise dependencies in the actor domain), it does not provide
guidance about
the structure of the temporal dependence.
We accommodate temporal dependence by expanding the model to allow the
random effects to be correlated over time.
We consider the following first order (Markov) auto-regressive
structure for the random effects:
%
%e2 ###
\begin{eqnarray}
\label{ModAREqn}
(
s_{i,t}, r_{i,t}
)' & = & \Phi_{sr}
(
s_{i,t-1}, r_{i,t-1}
 )' + \epsilon_{sr,t},
\nonumber\\
(
g_{i,j,t}, g_{j,i,t}
 )' & = & \Phi_{gg}
 (
g_{i,j,t-1}, g_{j,i,t-1}
 )' + \epsilon_{gg,t},\\
  \eqntext{\mbox{where }
\Phi_{sr} =
 \pmatrix{\displaystyle
\phi_s & \phi_{sr} \cr\displaystyle
\phi_{rs} & \phi_r
}
 ,\ \Phi_{gg} =
 \pmatrix{\displaystyle
\phi_g & \phi_{gg} \cr\displaystyle
\phi_{gg} & \phi_g
} ,}
\end{eqnarray}
and $\epsilon_{sr,t}$ and $\epsilon_{gg,t}$ are independent mean-zero
bivariate normal vectors with covariance matrices $\Gamma_{sr}$ and
$\Gamma_{gg}$:
%
%e3 ###
\begin{equation}
\label{ModStaCovEqn}
\Gamma_{sr} =
 \pmatrix{\displaystyle
\gamma^2_s & \gamma_{sr} \cr\displaystyle
\gamma_{sr} & \gamma^2_r
}
,  \qquad
\Gamma_{gg} =
 \pmatrix{\displaystyle
\gamma^2_g & \lambda_{gg}\gamma_{g}^2 \cr\displaystyle
\lambda_{gg}\gamma_{g}^2 & \gamma^2_g
}%
.
\end{equation}
The resulting covariance matrix for
the vector
$sr_i = ( s_{i,1},r_{i,1},\ldots , s_{i,T},r_{i,T} )' $
can be written as
\[
\Cov( sr_i) = \Sigma_{sr}=  \pmatrix{\displaystyle
\Sigma_{sr}(0) &\Sigma_{sr}(1) & \cdots & \Sigma_{sr}(T-1) \cr\displaystyle
\Sigma_{sr}(1)' &\Sigma_{sr}(0) & \cdots & \Sigma_{sr}(T-2) \cr\displaystyle
\vdots& \vdots & & \vdots \cr\displaystyle
\Sigma_{sr}(T-1)' &\Sigma_{sr}(T-2)' & \cdots & \Sigma_{sr}(0)
},
\]
where $\Sigma_{sr}(d)$ depends on $\Phi_{sr}$, $\Gamma_{sr}$ and the time
lag $d$.
The covariance matrix of the vector
$g_{[i,j]} = ( g_{i,j,1},g_{j,i,1},\ldots , g_{i,j,T},g_{j,i,T} )' $
has a similar block Toeplitz structure, which we write as
$\Cov(g_{[i,j]} ) = \Sigma_{gg} $, and is made up of the blocks
$\{ \Sigma_{gg}(0),\ldots , \Sigma_{gg}(T-1) \}$.
Putting the two sources of variation together,
Table~\ref{ModStaCovStr} outlines the set of potentially nonzero
covariances defined by the random effects model. The different $\sigma
$'s in Table~\ref{ModStaCovStr} replace their more general counterparts,
the $\xi$'s of Table
\ref{GenCovStr}.

%t2 ###
\begin{table}%[h]
\caption{Covariances based upon the stationary mixed effects model}
\label{ModStaCovStr}
\begin{tabular}{@{}llcc@{}}
\hline
&&  $\bolds{d=0}$ & $\bolds{d>0}$ \\
\hline
(a)& $\Cov(y_{i,j,t},y_{i,k,t+d})= $ &$\sigma^2_s$ & $\sigma
_{s,d}$ \\
[3pt]
(b)& $\Cov(y_{i,j,t},y_{k,j,t+d}) =$ & $\sigma^2_r$ & $\sigma
_{r,d}$ \\
[3pt]
\multirow{1}{10pt}[-6pt]{(c)}&
$\Cov(y_{i,j,t},y_{j,k,t+d}) =$&$\sigma_{sor}$&$\sigma_{rs,d}$ \\
&$\Cov(y_{i,j,t+d},y_{j,k,t}) =$&$\sigma_{{sor}}$&$\sigma_{sr,d}$
\\[3pt]
(d)&   $\Cov(y_{i,j,t},y_{i,j,t+d}) =$ & $\sigma^2_s + \sigma
^2_r + \sigma^2_g$ & $\sigma_{s,d} + \sigma_{r,d} + \sigma_{g,d}$\\
[3pt]
(e)&  $\Cov(y_{i,j,t},y_{j,i,t+d}) =$ & $\sigma_{gg} + 2\sigma
_{{sor}}$ & $\sigma_{gg,d} + \sigma_{sr,d} + \sigma_{rs,d}$ \\
\hline
\end{tabular}
\end{table}

Note that if we were to consider a static network, the covariances
given by~$\Sigma_{sr}(0)$ and $\Sigma_{gg}(0)$ would represent those
for the social relations models as outlined in \citet{WarKenSto79} and
\citet{GilSwa01}. As such, those models are submodels of the one
defined by equations (\ref{ModEqn})--(\ref{ModStaCovEqn}).

Through the use of a generalized linear model [\citet{McCNel}], the
mixed effects model for Gaussian longitudinal social relations data can
be extended to analyze relations that are not appropriately modeled by
a Gaussian distribution, such as binary responses or counts. This is
done by using the above model to describe a linear predictor $\theta
_{i,j,t}$ in a generalized linear model. This leads to the following
formulation:
\begin{eqnarray*}
E(y_{i,j,t}|\theta_{i,j,t}) &=& h(\theta_{i,j,t}), \\
\theta_{i,j,t} &=& x_{i,j,t}'\beta_t + s_{i,t} + r_{j,t} + g_{i,j,t}.
\end{eqnarray*}
Under the model, the $y_{i,j,t}$'s are conditionally independent
given the $\theta_{i,j,t}$'s, so that we have
\[
p(y | \theta) = \prod_{i=1}^{A-1} \prod_{j=i+1}^A \prod_{t=1}^T
p(y_{i,j,t}|\theta_{i,j,t})p(y_{j,i,t}|\theta_{j,i,t}).
\]
 The covariance structure here is approximately that of the
Gaussian model multiplied by a factor depending on the link function
$h$ [\citet{Hofbi}; \citet{Wes07}],
indicating that the second order dependence outlined by Figure~\ref
{DepStr} is still captured:
\[
\Cov( y_{i_1, j_1, t_1}, y_{i_2, j_2, t_2})
\approx
\Cov(\theta_{i_1, j_1, t_1}, \theta_{i_2, j_2, t_2}) \times
h'(x_{i_1, j_1,t_1}' \beta_{t_1})h'(x_{i_2, j_2,t_2}' \beta_{t_2}).
\]
%
%%%%%%%%%%%%%%%%%%%%%%%%%%%%%%%%%%

%s4 ###
\section{Parameter estimation}\label{sec4}
Estimation of model parameters is most easily done in the context of Bayesian
inference.
In this section we present a~general Markov chain Monte Carlo (MCMC)
algorithm for continuous data which are modeled as Gaussian, and binary
data which are modeled through a~particular probit formulation based on
the work of \citet{AlbChi93} and \citet{ChiGre98}.

%%%%%%%%%%%%%%%%%%%%%%%%%%%%%%%%%%%%
%s4.1 ###
\subsection{Gaussian mixed effects model}\label{sec4.1}

The model fully defined by equations~(\ref{ModEqn})--(\ref
{ModStaCovEqn}) has the following parameters $\Theta$ that need to be
estimated: $\Theta= \{ (\beta_t; t=1,\ldots , T), (\phi_s, \phi
_{sr}, \phi_{rs}, \phi_r), (\phi_g, \phi_{gg}), (\gamma^2_s, \gamma
^2_r, \gamma_{sr}),   (\gamma^2_g, \lambda_{gg}), ( s_{i,t},
r_{i,t}; \break  i=1, \ldots ,A;t=1,\ldots ,T) \}.$ A
Bayesian analysis is conducted by examining the joint distribution of
the parameters in $\Theta$ given the data $y$:
%
%
%e4 ###
\begin{eqnarray}
\label{PostDistAR}
P( \Theta| y) &\propto& \prod_{i=1}^{A-1} \prod_{j=i+1}^A \operatorname
{dmvn}\bigl(y_{[i,j]} | \eta_{[i,j]}+ sr_i + rs_j, \Sigma_{gg}\bigr)\nonumber
\\[-10pt]
\\[-10pt]
&& {}\times
\prod_{i=1}^A \operatorname{dmvn}(sr_i| 0, \Sigma_{sr})
\times P(\beta) P(\Phi_{gg}) P(\Gamma_{gg}) P(\Phi_{sr}) P(\Gamma
_{sr}),
\nonumber
\end{eqnarray}
 where ``dmvn'' stands for a multivariate normal density
function and
\begin{eqnarray*}
y_{[i,j],t} &=& ( y_{i,j,t}, y_{j,i,t})', \qquad
y_{[i,j]}=
\bigl(y_{[i,j],1}',
\ldots, y_{[i,j],T}' \bigr)' , \\
\eta_{[i,j],t} &=& ( \beta_t'x_{i,j,t}, \beta_t'x_{j,i,t})', \qquad
\eta_{[i,j]}=
\bigl(\eta_{[i,j],1}',
\ldots, \eta_{[i,j],T}' \bigr)' , \\
sr_{i,t} &=& ( s_{i,t}, r_{i,t} )', \qquad
sr_{i} = ( sr_{i,1}', \ldots, sr_{i,T}' )' ,\\
rs_{i,t} &=& ( r_{i,t}, s_{i,t} )', \qquad
rs_{i} = ( rs_{i,1}', \ldots, rs_{i,T}' )'.
\end{eqnarray*}
The first double product of equation (\ref{PostDistAR}) is the density of
the data given the sender--receiver random effects, the next product is
the sampling distribution of the random effects, and the remaining
terms are the priors for the model. We use the following
semi-conjugate priors for $\beta, \Phi_{sr}, \Phi_{gg}$, and $\Gamma_{sr}$:
\begin{eqnarray*}
\beta= (\beta_1',\ldots,\beta_T')'
& \sim& \operatorname{mvn}(M_{\beta},
V_{\beta}), \\
(\phi_s, \phi_{sr}, \phi_{rs}, \phi_r)'
& \sim& \operatorname{mvn}(M_{\Phi
_{sr}}, V_{\Phi_{sr}}) \mathbb{I}(\Phi_{sr} \in\mathcal{S}), \\
(\phi_g, \phi_{gg})'
& \sim& \operatorname{mvn}(M_{\Phi_{gg}}, V_{\Phi
_{gg}}) \mathbb{I}(\Phi_{gg} \in\mathcal{S}), \\
\Gamma_{sr}
& \sim& \operatorname{inverse\mbox{-}Wishart}(v_{sr}, S_{sr}^{-1}).
\end{eqnarray*}
The $\phi$-parameters are constrained to ensure that the temporal
processes for the sender--receiver effects and the residual error terms
produce a stationary process~$\mathcal{S}$ [\citet{Rei97}].
Such a constraint allows the fixed-effects and covariance parameters
to represent means and variances of the observed data over the observed
time period.
For an $\operatorname{AR}(1)$ model, the constraint is satisfied if the absolute value
of eigenvalues for the $\Phi$'s are less than 1.

A conjugate prior for the Toeplitz matrix $\Gamma_{gg}$ can be obtained
by considering a transformation described by \citet{Won82}. In order to
apply this approach to our problem, we consider the following bivariate
innovations to obtain independent bivariate distributions:
\begin{eqnarray*}
( \tilde{g}_{i,j,t}, \tilde{g}_{j,i,t})'
&=& ( g_{i,j,t}, g_{j,i,t} )'
- \Phi_{gg} ( g_{i,j,t-1}, g_{j,i,t-1} )' \\
&\sim&
\operatorname{mvn}(0, \Gamma_{gg}).
\end{eqnarray*}
Now using the property of bivariate normal distributions, we can create
two independent vectors: $a_{i,j,t} = \tilde{g}_{i,j,t} + \tilde
{g}_{j,i,t}$ and $b_{i,j,t} = \tilde{g}_{i,j,t} - \tilde{g}_{j,i,t}$,
where $a_{i,j,t} \sim\operatorname{normal} (0, \sigma_{a}^2)$ and
$b_{i,j,t} \sim\operatorname{normal} (0, \sigma_{b}^2)$. We use
inverse-gamma priors for $\sigma_{a}^2$ and $\sigma_{b}^2$:
$\sigma_{a}^2 \sim$ $\operatorname{inverse\mbox{-}gamma}(\alpha_{a}, \delta_{a})$,
$\sigma_{b}^2 \sim$ $\operatorname{inverse\mbox{-}gamma}(\alpha_{b}, \delta_{b}).$
The matrix $\Gamma_{gg}$ can be constructed as $\gamma_{g}^2 = (\sigma
_{a}^2 + \sigma_{b}^2)/4$ and $\lambda_{gg} = (\sigma_{a}^2 - \sigma
_{b}^2)/\break(\sigma_{a}^2 + \sigma_{b}^2)$.

Based on this class of prior distributions,
a Markov chain Monte Carlo approximation to the joint posterior
distribution may be obtained via Gibbs sampling for the $\beta$'s and
the sender--receiver effects, with
a Metropolis--Has\-tings update for $\Phi_{sr}$, $\Phi_{gg}$, $\Gamma
_{sr}$, and $\Gamma_{gg}$. However, the Metropolis--Hastings updates
are based on their full conditional distributions.
For example, consider that the full conditional distribution of $\Phi
_{sr}$ is given by
\begin{subequation}
%
%e7 ###
%e6 ###
%e5 ###
\begin{eqnarray}
\label{conprop1}
P( \Phi_{sr} | \cdot) &\propto& \prod_{i=1}^A\operatorname{dmvn} (sr_{i,1} |
0, \Sigma_{sr}(0) )\\
\label{conprop2}
&& {}\times \prod_{i=1}^A \prod_{t=2}^T\operatorname{dmvn} (sr_{i,t} | \Phi_{sr}
sr_{i,t-1}, \Gamma_{sr} ) \\
\label{conprop3}
&&{} \times\operatorname{dmvn}(\Phi_{sr} | M_{\Phi_{sr}}, V_{\Phi_{sr}}) \mathbb
{I}(\Phi_{sr} \in\mathcal{S}).
\end{eqnarray}
\end{subequation}
If we were to ignore the first product [equation (\ref{conprop1})] and
the stationarity constraint, the expression above would be proportional
to a multivariate normal distribution. Since most of the information
about $\Phi_{sr}$ is contained in equation (\ref{conprop2}) and (\ref
{conprop3}), the full conditional distribution of $\Phi_{sr}$ will be
close to this multivariate normal distribution. We use this
approximation to the full conditional distribution as a proposal
distribution, but make the necessary correction in the acceptance
probability via the Metropolis--Hastings algorithm. We use a similar
Metropolis--Hastings proposal for updating $\Phi_{gg}$.
Further details about the MCMC algorithm, including information for
updating $\Gamma_{sr}$ and $\Gamma_{gg}$, can be found in the \hyperref[appm]{Appendix}.

%%%%%%%%%%%%%%%%%%%%%%%%%%%%%%%%%%%%%
%s4.2 ###
\subsection{Probit mixed effects model}\label{sec4.2}
In order to model data that are not approximately Gaussian, such as
binary data, we move the Gaussian structure to a secondary level in the
hierarchical model leading to the following formulation:
\begin{eqnarray*}
y_{i,j,t}
 &\sim& p(y|\theta_{i,j,t}),  \\
\theta_{i,j,t}
&=& x_{i,j,t}' \beta_t + s_{i,t} + r_{j,t} + g_{i,j,t},
\end{eqnarray*}
 where $p(y|\theta)$ represents the probability distribution
of the response. For example, a probit model for binary data can be
obtained by
setting $p(y|\theta) = \Phi(\theta)^{y} [1-\Phi(\theta)]^{1-y}$.
For the probit model, we specify the covariance of the sender--receiver
effects $\Sigma_{sr}$ as before based upon the parameters $\Phi_{sr}$
and~$\Gamma_{sr}$. However, as noted in \citet{AlbChi93}, the variance
parameter~$\sigma^2_{gg}$ in covariance matrix $\Sigma_{gg}$ is not
identifiable. For ease of interpretation, we will set $\sigma^2_{gg}$
to be equal to one so that $\Sigma_{gg}$ is a correlation matrix. In
doing this, additional constraints are placed on $\Phi_{gg}$ and $\Gamma
_{gg}$. Consider the following Yule--Walker equations for a first order
auto-regressive process:
\[
\label{YW}
\Cov\bigl(g_{[i,j],t}, g_{[i,j],t+d}\bigr) = \Sigma_{gg}(d) =
\cases{\displaystyle
\Sigma_{gg}(d-1) \Phi_{gg}' + \Gamma_{gg}, &\quad   if    $d=0$, \cr\displaystyle
\Sigma_{gg}(d-1) \Phi_{gg}', &\quad   if    $d>0$.
}
\]
Since $\Sigma_{gg}$ is a correlation matrix, $\Sigma_{gg}(0)$ is also a
correlation matrix, with a correlation coefficient $\rho_{gg}$. Solving
the Yule--Walker equations in terms of~$\Sigma_{gg}(0),$ we have
\[
\Sigma_{gg}(0) = \Phi_{gg} \Sigma_{gg}(0) \Phi_{gg} + \Gamma_{gg}.
\]
Writing this out in terms of the individual parameters results in
\begin{eqnarray*}
 \pmatrix{\displaystyle
1 & \rho_{gg} \cr\displaystyle
\rho_{gg} & 1
}
&=&
\pmatrix{\displaystyle
\phi_{g} & \phi_{gg} \cr\displaystyle
\phi_{gg} & \phi_{g}
}
 \pmatrix{\displaystyle
1 & \rho_{gg} \cr\displaystyle
\rho_{gg} & 1
} \\
&&{}\times \pmatrix{\displaystyle
\phi_{g} & \phi_{gg} \cr\displaystyle
\phi_{gg} & \phi_{g}
}
+
 \pmatrix{\displaystyle
\gamma^2_{g} & \gamma_{gg} \cr\displaystyle
\gamma_{gg} & \gamma^2_{g}
}.
\end{eqnarray*}
Now we solve for $\gamma^2_{g}$ and $\gamma_{gg}$ in terms of $\phi
_{g}, \phi_{gg}$ and $\rho_{gg}$ to get%
\begin{eqnarray*}
\gamma^2_{g} &=& 1 - \phi_{g}^2 - \phi_{gg}^2 - 2 \rho_{gg} \phi_{g}
\phi_{gg},\\
\gamma_{gg} &=& \rho_{gg} -2 \phi_{g} \phi_{gg} - \rho_{gg} \phi_{g}^2
-\rho_{gg} \phi_{gg}^2.
\end{eqnarray*}
If we consider a Bayesian estimation algorithm, we can propose values
of~$\phi_{g}$, $\phi_{gg}$ and $\rho_{gg}$ such that $\Gamma_{gg}$ is a
proper covariance matrix and it is guaranteed that $\Sigma_{gg}$ will
be a correlation matrix.

The joint density of the parameters conditional on the data $y$ is
proportional to
%
%
%e8 ###
\begin{eqnarray}
\label{PostDistGLMAR}
p( \Theta| y) &\propto&\prod_{i=1}^{A-1} \prod_{j=i+1}^A P\bigl(y_{[i,j]}
| \theta_{[i,j]}\bigr) \times
\operatorname{dmvn}\bigl(\theta_{[i,j]} | \eta_{[i,j]} + sr_i + rs_j, \Sigma
_{gg}\bigr) \nonumber
\\[-8pt]
\\[-8pt]
&&{}\times \prod_{i=1}^A \operatorname{dmvn}(sr_i| 0, \Sigma_{sr})
\times P(\beta) P(\Phi_{gg}) P(\rho_{gg}) P(\Phi_{sr}) P(\Gamma_{sr}).
\nonumber
\end{eqnarray}
Because of the nonidentifiability and reparameterization of $\Gamma
_{gg}$ discussed above, we impose constraints on $\Gamma_{gg}$ and $\Phi
_{gg}$ via the following priors:
\begin{eqnarray*}
\rho_{gg} &\sim& \operatorname{normal}(M_{\rho_{gg}}, V_{\rho_{gg}}) \mathbb
{I}(-1 \leq \rho_{gg} \leq1) \mathbb{I}(\Gamma_{gg} \mbox{ is
positive definite}), \\
(\phi_g, \phi_{gg})'
& \sim& \operatorname{mvn}(M_{\Phi_{gg}}, V_{\Phi
_{gg}})\mathbb{I}(\Phi_{gg} \in\mathcal{S}) \mathbb{I}(\Gamma_{gg}
\mbox{ is positive definite}).
\end{eqnarray*}
To estimate the model parameters, the MCMC algorithm presented in
Section~\ref{sec4.1} is modified in two ways: (1) $\rho_{gg}$ is now explicitly
updated, and (2) the latent response $\theta_{i,j,t}$ must also be
updated. For most GLMs a Metropolis--Hastings step is required to
update the latent response. However, the probit model allows for a
Gibbs sampling procedure based upon the work of \citet{AlbChi93} and
\citet{ChiGre98}. The Gibbs sampling procedure for each $(i,j)$ and
$(j,i)$ pair at times $t=1,\ldots,T$ proceeds by sampling the
conditional distribution for each $\theta_{i,j,t}$, based on a
truncated normal distribution: The truncation is to the left of zero if
$y_{i,j,t}=0$ and to the right of zero if $y_{i,j,t}=1$. Further
details on the MCMC algorithm can be found in the \hyperref[appm]{Appendix}.

%%%%%%%%%%%%%%%%%%%%%%%%%%%%%%%%%%%%
%s5 ###
\section{International trade}\label{sec5}
In this section we apply the methodology to the study of yearly
international trade between 58 countries from 1981--2000.\footnote{A
list of countries (including their three-letter ISO codes) used in this
analysis can be found in the \hyperref[appm]{Appendix}. Additionally, the data and some
of the R code used to fit the model are available as supplementary
material [\citet{WesHofLSNSup}].}
A~commonly used model for international trade is the gravity model
[\citet{Tin62}] which, based
on Newton's law of gravity, posits that the force of trade between two
countries is proportional to the product of their economic ``masses''
divided by the distance between them (raised to some power).
Taking logs, a formulation of the gravity model
in the context of longitudinal trade
is given by
\[
\ln \mathrm{Trade}_{i,j,t} =\beta_{0,t}+ \beta_{1,t} \ln \mathrm{GDP}_{i,t}+ \beta_{2,t}\ln \mathrm{GDP}_{j,t} + \beta_{3,t} \ln \mathrm{D}_{i,j,t} + \epsilon_{i,j,t},
\]
where $\mathrm{Trade}_{i,j,t}$ is the trade between two countries at time $t$,
$\mathrm{D}_{i,j}$ the geographic distance between them, and
$\mathrm{GDP}_{i,t}$ and $\mathrm{GDP}_{j,t}$ denote their gross domestic
products at time $t$.\footnote{As opposed to \citet{WarHof07} and \citet
{Wes07}, real values for GDP and the level of trade were used in this
paper. The reason the other works used nominal values was to avoid
modeling the inflation rate for out of sample prediction. An inflator
using the CPI-All Urban Consumers data was calculated to set the
amounts into real values based on the year 2000. The CPI data can be
obtained from the following:
\href{http://www.bls.gov/data/home.htm}{http://}
\href{http://www.bls.gov/data/home.htm}{www.bls.gov/data/home.htm}.
Note: this CPI data is used in the BLS inflation calculator:
\href{http://data.bls.gov/cgi-bin/cpicalc.pl}{http://data.bls.gov/cgi-bin/cpicalc.pl}.}

Over the past forty years the gravity model of bilateral trade has
become a benchmark for several reasons: (1) A gravity model can
typically explain about one-half the variation in bilateral
international commerce [\citet{WarHof07}]; (2) The gravity model can be
derived from first principles of economic theory [\citet{And79}]; (3)
The linear formulation of the model is easy to work with empirically
and readily accommodates other factors that might affect trade flows.

Following \citet{WarHof07}, we will consider two other factors for this
analysis: the polity of a nation and whether pairs of nations
cooperated in militarized interstate disputes.
Polity, denoted by Pol, measures a nation's level of democracy, and
ranges from 0 for highly authoritarian regimes to 20 for highly
democratic ones. Cooperation in conflict, denoted by CC, measures
active military cooperation. If the pair cooperated on a~particular
dispute, it receives a value of $+$1. However, if the two countries were
on opposite sides of a dispute, a value of $-$1 is recorded. If there was
more than one dispute in a single year involving the same pair, then
the pair's scores are summed over all disputes in that year. It should
be noted that all of the covariates except distance are changing over
time.\footnote{For further discussion of the data used in this paper,
we refer the reader to \citet{WarHof07}.} This leads to the following
model, which is motivated by the gravity model, additional covariates
of interest and the longitudinal network structure:
\begin{eqnarray*}
\ln \mathrm{Trade}_{i,j,t}
&=& \beta_{0,t} + \beta_{1,t} \ln \mathrm{GDP}_{i,t} + \beta_{2,t} \ln \mathrm{GDP}_{j,t} +
\beta_{3,t}\ln \mathrm{D}_{i,j,t} \\
&&{} +\beta_{4,t} \mathrm{Pol}_{i,t} + \beta_{5,t} \mathrm{Pol}_{j,t} +
\beta_{6,t} \mathrm{CC}_{i,j,t} + \beta_{7,t} \mathrm{Pol}_{i,t} \times
\mathrm{Pol}_{j,t}  \\
&&{}+ s_{i,t} + r_{j,t} + g_{i,j,t}
\end{eqnarray*}
 with the following diffuse priors:
\begin{eqnarray*}
\beta_t & \sim& \operatorname{mvn}(0, 100 \times\mathrm{I}), \\
(\phi_s, \phi_{sr}, \phi_{rs}, \phi_r)' & \sim& \operatorname{mvn}(0, 100
\times\mathrm{I}) \mathbb{I}(\Phi_{sr} \in\mathcal{S}), \\
(\phi_g, \phi_{gg})' & \sim& \operatorname{mvn}(0, 100 \times\mathrm{I})
\mathbb{I}(\Phi_{gg} \in\mathcal{S}), \\
\Gamma_{sr} & \sim& \operatorname{inverse\mbox{-}Wishart}(4, \mathrm{I}),\\
\sigma_{a}^2 & \sim& \operatorname{inverse\mbox{-}gamma} (1,1),\\
\sigma_{b}^2 & \sim& \operatorname{inverse\mbox{-}gamma} (1,1).
\end{eqnarray*}
Initially we implemented the MCMC algorithm outlined in Section~\ref{sec4.1},
however, we found the Markov chain to be very ``sticky.'' This result may
have occurred since the semi-conjugate Gibbs proposals are similar to
an independence proposal. In this case the distribution of the proposal
should be close to the respective posterior distribution but should be
``fatter'' in the tails to prevent ``stickiness'' [\citet{GivHoe05}]. This
would suggest that we should increase the variance of the
semi-conjugate Gibbs proposals to increase the rate of mixing. However,
the posterior distribution of $\Phi_{sr}$ is near the boundary for
stationary processes,
and increasing the variance of the proposals may lead to more
unaccepted proposed values. Therefore, to safeguard against poor mixing
of the chain, we randomly alternated between using (1) semi-conjugate
Gibbs proposals (without an increased variance), and (2) random walk
proposals around the current values of the parameters ($\Phi_{sr},
\Sigma_{sr}, \Phi_{gg}, \Sigma_{gg}$). A Markov chain of 55,000
iterations was
generated, the first 10,000 of which were dropped to allow convergence to
the stationary distribution.
Parameter values were saved every 20th scan, resulting in
2,250 samples
with which to approximate the joint posterior distribution.
%
%%%%%%%%%%%%%%%%%%%%%%%%%%%%%%%%%%%%%%%%%%%%%%
%s5.1 ###
\subsection{Results}\label{sec5.1}

\begin{figure}

\includegraphics{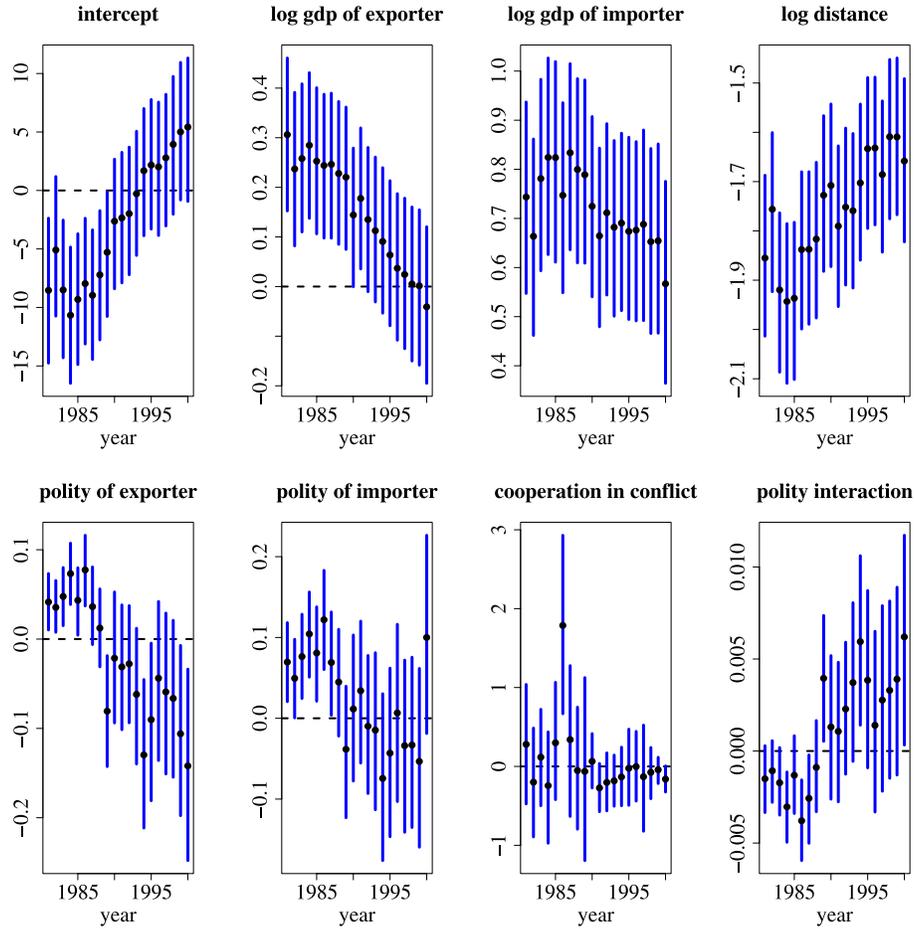}

\caption{95\% credible intervals of the covariate coefficients over
time for the gravity model.}
\label{PDBeta}
\end{figure}

The 95\% posterior credible intervals (blue bars) and their medians
(black dots) for the $\beta$'s are in Figure \ref{PDBeta}. Let us first
consider the panels on the top row, excluding the intercept. The
posterior distributions of the coefficients in the gravity model have
several features: (1) In general, the credible intervals of the
coefficients for the $\ln \mathrm{GDP}$ of the exporter are shifting downward over
the period. Additionally, these intervals contain zero from 1994 to
2000, heuristically suggesting that this covariate is becoming a~less
important correlate of bilateral trade flows.
(2) The coefficients for the $\ln \mathrm{GDP}$ of the importer over the period are
all positive, suggesting that the economic size of the importer is an
important factor in bilateral trade flows. (3) As might be expected,
over the twenty-year period the medians of the coefficients for
distance are generally decreasing. An intuitive explanation is that the
transportation of goods and services has become more efficient over the period.

%f2 ###

The four panels on the bottom row of Figure \ref{PDBeta} are the
results for the additional predictors of trade beyond the gravity
model. There appears to be a general decline in the coefficients for
the main effects of polity of the exporter and importer over the
period, with a notable exception for the latter in the year 2000 (the
95\% credible interval still contains zero). However, there appears to
be a rising trend in the coefficients of polity interaction ($\beta
_{7,t}$) over the period. The trend suggests that trade between
democratic countries is increasing faster than the average. Finally,
for the polity coefficients in general we see that our estimate is
becoming more uncertain over time, as the credible intervals are
widening over the period. A plausible explanation for this phenomenon
is that the countries under study are becoming more democratic, thus,
there is less variation in the polity covariate. The sample mean and
variance of the polity score for 1980 are 3.62 and 56.66, respectively,
while in 2000 they are 7.43 and 20.56. Based upon the model, whether
two nations cooperate in conflicts is not indicative of the level of
trade between them, except for the notable case of 1986, where
bilateral trade is positively correlated with military cooperation.

%f3 ###
\begin{figure}

\includegraphics{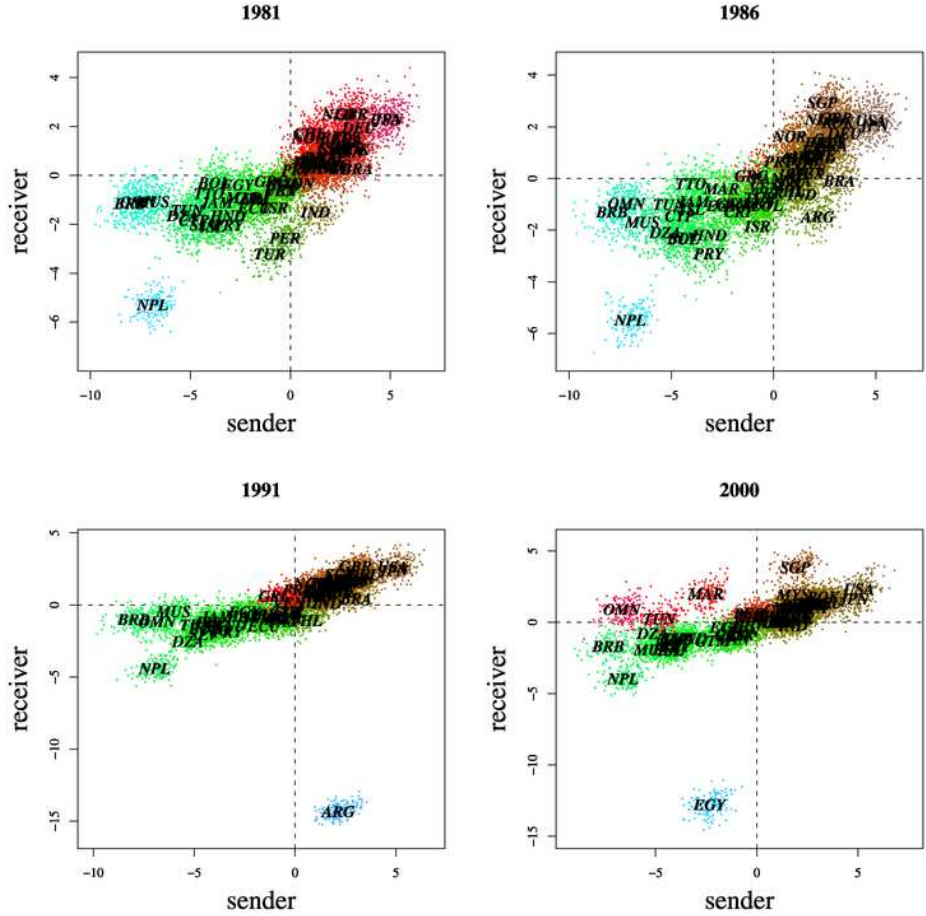}
\vspace*{-3pt}
\caption{Posterior distributions of the sender--receiver effects over
time for the gravity model.}
\label{PDBetaSR}
\vspace*{-3pt}
\end{figure}

We now examine the posterior distributions of the country-specific
sender and receiver random effects.
These effects describe the average deviations of a country's export and
import levels from those that would be predicted by the regression
model alone.
In Figure \ref{PDBetaSR} the colored dots
are a random sample of 150 values from the bivariate posterior
distribution of the sending and receiving effects for each country, and
the country labels are located at the posterior means. Countries that
are close to each other, based on their posterior mean, are similar in
color. As might be expected, we see in each plot that there exists a
strong positive relationship between exporting (sending) and importing
(receiving) and that the relative positions of the nations change only
slightly over the four years shown in the figure.
This strong positive relationship suggests a possible model simplification
for these data in which the sender and receiver effects are co-linear, although
such a~model reduction may not be appropriate for other data sets.

%
%
%t3 ###
\begin{table}
\tabcolsep=0pt
\tablewidth=224pt
\caption{$\Sigma(0)_{sr}$ and $\Sigma(0)_{gg}$ parameter estimates for
the gravity model}
\label{BetaGamma}
\vspace*{-3pt}
\begin{tabular*}{\tablewidth}{@{\extracolsep{\fill}}lcccc@{}}
%d{2.3}d{2.3}d{2.3}@{}}
\hline
\multicolumn{1}{c}{\textbf{Parameter}} &
\multicolumn{1}{c}{\textbf{Markov chain}} &
\multicolumn{1}{c}{\textbf{2.5\%}} &
\multicolumn{1}{c}{\textbf{Median}} &
\multicolumn{1}{c@{}}{\textbf{97.5\%}}
\\
\hline
$\sigma^2_s$ &
\includegraphics{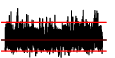} &
$\textcolor{red}{\phantom{1}6.777}$ &
$\textcolor{black}{\phantom{1}9.841}$ &
$\textcolor{red}{14.733}$
\\
$\sigma_{{sor}}$ &
\includegraphics{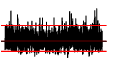} &
$\textcolor{red}{\phantom{1}2.429}$ &
$\textcolor{black}{\phantom{1}3.665}$ &
$\textcolor{red}{\phantom{1}5.591}$
\\
$\rho_{sr}$ &
\includegraphics{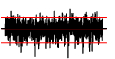} &
$\textcolor{red}{\phantom{1}0.597}$ &
$\textcolor{black}{\phantom{1}0.705}$ &
$\textcolor{red}{\phantom{1}0.790}$
\\
$\sigma^2_r$ &
\includegraphics{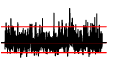}  &
$\textcolor{red}{\phantom{1}2.129}$ &
$\textcolor{black}{\phantom{1}2.787}$ &
$\textcolor{red}{\phantom{1}3.829}$
\\
$\sigma^2_g$ &
\includegraphics{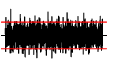} &
$\textcolor{red}{10.089}$ &
$\textcolor{black}{10.292}$ &
$\textcolor{red}{10.496}$
\\
$\sigma_{gg}$ &
\includegraphics{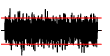} &
$\textcolor{red}{\phantom{1}3.113}$ &
$\textcolor{black}{\phantom{1}3.327}$ &
$\textcolor{red}{\phantom{1}3.523}$
\\
$\rho_{gg}$ &
\includegraphics{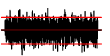} &
$\textcolor{red}{\phantom{1}0.307}$ &
$\textcolor{black}{\phantom{1}0.323}$ &
$\textcolor{red}{\phantom{1}0.338}$
\\
\hline
\end{tabular*}
\vspace*{-6pt}
\end{table}

A closer examination of the plots reveals that the United States (USA),
Germany (DEU), Japan (JPN) and the United Kingdom (GBR) are located at
the top right corner for most of these plots, and thus are considered
some of the most active nations in the network, even after accounting
for their covariate information. On the other hand, nations such as
Nepal (NPL), Oman (OMN), Barbados (BRB) and Mauritius (MUS) are among
the least active, based on their location in the plots. Over the
period, the rise of East Asian countries through trade is exemplified
by the movement of Singapore (SGP) on the receiving axis---the 95\%
credible interval of Singapore's receiving position in 2000 minus its
receiving position in 1981 is (1.449, 3.967). Finally, note the dip in
imports to Argentina (ARG) in 1991 and Egypt (EGY) in 2000. In each
situation, the value of imports from all countries in the data is zero.
It is unlikely that there were no imports for either country for those
years. A plausible explanation for the imports to Argentina being
``zeroed-out'' might be due to a currency reform that the country
undertook in 1991. As for the Egyptian case, around the year 2000 there
was not a period of financial instability, suggesting that the zero
imports are an aberration in the data.
We note that, by allowing for time and country-specific importer and
exporter effects, our estimates of the regression coefficients will be
fairly robust to such outliers.

The assumption of a stationary covariance structure
allows us to interpret the
the marginal covariances $\Sigma(0)_{sr}$ and $\Sigma(0)_{gg}$
as across-year average covariances. Using the posterior samples from
$\Phi_{sr}, \Gamma_{sr},\Phi_{gg}$ and $\Gamma_{gg}$, the empirical
posterior distributions for $\Sigma(0)_{sr}$ and $\Sigma(0)_{gg}$ can
be computed. The results are in Table \ref{BetaGamma}, which presents
the trace plots of the Markov chains along with the 95\% credible
intervals and posterior medians. Notice that the medians of the
posterior distributions for $\sigma^2_s$ and $\sigma^2_r$ coincide with
the spread of the posteriors of the sender--receiver estimates for the
nations (Figure \ref{PDBetaSR}). Also from the table we see that: (1)
the median posterior correlation $\rho_{sr}$ between the sending and
receiving effects is 0.705, and (2) the median posterior residual
correlation $\rho_{gg}$ within a pair of nations is 0.323. The latter
suggests a modest degree of reciprocity among pairs of actors in the
network at a given point in time.

%t4 ###
\begin{table}%[h]
\caption{$\Phi_{sr}$ and $\Phi_{gg}$ parameter estimates for the
gravity model}
\label{BetaARTrade}
\vspace*{-3pt}
\begin{tabular}{@{}lcccc@{}}
\hline
\textbf{Parameter} &
\textbf{Markov chain} &
\multicolumn{1}{c}{\textbf{2.5\%}} &
\multicolumn{1}{c}{\textbf{Median}}  &
\multicolumn{1}{c@{}}{\textbf{97.5\%}}
\\
\hline
$\phi_s$ &
\includegraphics{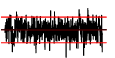} &
$\textcolor{red}{\phantom{-}0.991}$ &
 $\textcolor{black}{0.997}$ &
 $\textcolor{red}{1.002}$
 \\
$\phi_{sr}$ &
\includegraphics{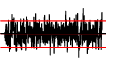} &
$\textcolor{red}{-0.010}$ &
$\textcolor{black}{0.005}$ &
$\textcolor{red}{0.019}$
\\
$\phi_{rs}$ &
\includegraphics{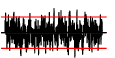} &
$\textcolor{red}{\phantom{-}0.121}$ &
 $\textcolor{black}{0.161}$ &
 $\textcolor{red}{0.201}$ \\
$\phi_r$ &
\includegraphics{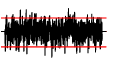} &
$\textcolor{red}{\phantom{-}0.505}$ &
 $\textcolor{black}{0.572}$ &
 $\textcolor{red}{0.632}$ \\
$\phi_g$ &
\includegraphics{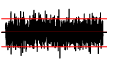} &
$\textcolor{red}{\phantom{-}0.665}$ &
 $\textcolor{black}{0.670}$ &
 $\textcolor{red}{0.676}$ \\
$\phi_{gg}$ &
\includegraphics{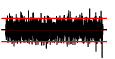} &
$\textcolor{red}{\phantom{-}0.100}$ &
 $\textcolor{black}{0.106}$ &
 $\textcolor{red}{0.111}$ \\
\hline
\end{tabular}
\vspace*{-6pt}
\end{table}

We also examine the auto-regressive coefficients to see what effect the
previous year has on exports, imports and reciprocity for the current
year. From Table \ref{BetaARTrade}, the medians of the posterior
distributions of $\phi_s$ and $\phi_{sr}$ are 0.997 and 0.005,
respectively. This suggests that the level of exports this year is
highly dependent on level of exports from the previous year but perhaps
not dependent on imports from the previous year. Comparatively, the
medians of the posterior distribution for $\phi_r$ and $\phi_{rs}$ are
0.572 and 0.161, respectively. That is, the level of imports this year
is fairly dependent on imports from the previous year and somewhat
dependent on exports from the previous year, indicating a possible
effect of increased purchasing power after a year of high exports.
Since the median of the posterior distribution of~$\phi_{gg}$ is 0.106,
we see that a relatively small amount of positive reciprocity in a
given year can be explained by the level of reciprocity in the previous
year.\looseness=-1

%%%%%%%%%%%%%%%%%%%%%%%%%%%%%%%%%%%%%
%s5.2 ###
\subsection{Out-of-sample prediction}\label{sec5.2}

In order to investigate the possibility that we are overfitting the
data, we randomly deleted 25\% of the responses, amounting to 16,120
cases, and compared the out-of-sample predictions for the LSR model
with covariates (M1) against four submodels (M2--M5). The first
submodel (M2) used the LSR structure but did not use any covariate
information ($y_{i,j,t} = \mu_t + s_{i,t} + r_{j,t} + g_{i,j,t}$). The
rest of the submodels considered (M3--M5) used covariate information
along with
either only network dependence, only temporal dependence, or neither
dependence structure:
\begin{itemize}[(M5)]
\item[(M3)] Social Relations Model:
\begin{eqnarray*}
\ln \mathrm{Trade}_{i,j,t}
&=& x'_{i,j,t} \beta_t + s_{i,t} + r_{j,t} +
g_{i,j,t}, \\
(
s_{i,t},
r_{i,t}
 )' & \sim& \operatorname{mvn} \left[ 0,  \pmatrix{\displaystyle
\gamma_s^2 & \gamma_{{sor}} \cr\displaystyle
\gamma_{{sor}} & \gamma_r^2
}  \right], \\
 (
g_{i,j,t},
g_{j,i,t}
 )' & \sim& \operatorname{mvn} \left[ 0,  \pmatrix{\displaystyle
\gamma_g^2 & \gamma_{gg} \cr\displaystyle
\gamma_{gg} & \gamma_g^2
}  \right].
\end{eqnarray*}
\item[(M4)] AR(1) Model:
\begin{eqnarray*}
\ln \mathrm{Trade}_{i,j,t}
&=& x'_{i,j,t} \beta_t + g_{i,j,t}, \\
g_{i,j,t}
&=& \phi_g g_{i,j,t-1} + \epsilon_{i,j,t};   \qquad    \epsilon
_{i,j,t} \sim\operatorname{normal} (0, \gamma^2).
\end{eqnarray*}
\item[(M5)] Standard Regression Model:\vspace*{5pt}
\begin{eqnarray*}
\ln \mathrm{Trade}_{i,j,t} &=& x'_{i,j,t} \beta_t + \epsilon_{i,j,t};
\qquad
   \epsilon_{i,j,t} \sim\operatorname{normal} (0, \gamma^2).\vspace*{5pt}
\end{eqnarray*}
\end{itemize}
%t5 ###
\begin{table}%[htdp]
\caption{Mean-squared-errors for LSR and submodels}
\label{MSE}
\vspace*{3pt}
\begin{tabular}{@{}lccd{2.3}@{}}
\hline
\textbf{Model} & \textbf{Temporal dep.} & \textbf{Network dep.} & \multicolumn{1}{c@{}}{\textbf{MSE}} \\
\hline
(M1) LSR Cov & yes & yes &4.665 \\
(M2) LSR mean & yes & yes & 4.681 \\
(M4) AR(1) & yes & no & 5.554 \\
(M3) Social relations & no & yes & 9.932 \\
(M5) Standard regression & no & no & 14.101\\
\hline
\end{tabular}
\vspace*{6pt}
\end{table}
%

%f4 ###
\begin{figure}

\includegraphics{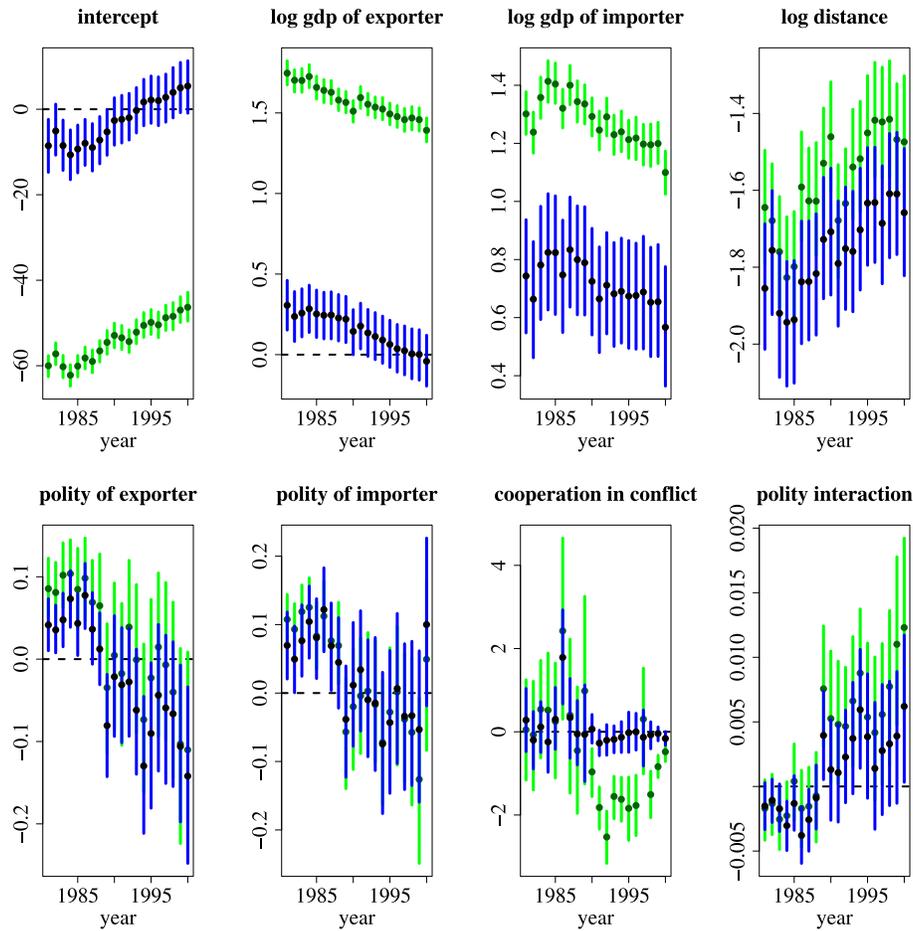}

\caption{Posterior 95\% credible intervals for the LSR (blue) and
standard regression model (green).}
\label{BetaLsnReg}
\end{figure}
For each of the five models, we used the median of the posterior of the
missing values as our predictor and compared the overall predictions
using the mean squared error score. Table \ref{MSE} presents these
scores for the LSR model with covariates and the four submodels. From
the ranking, the LSR model with covariates has the best performance,
suggesting that we may not be overfitting the data. Interestingly, the
next best model is the LSR mean model and is just slightly worse than
M1, suggesting that the covariates add little to the predictive
performance after the network and temporal dependence structures are
taken into account. The fact that the AR(1) model is next and performs
better than the Social Relations model suggests that there are strong
temporal dependencies in the data and these dependencies may be more
critical than capturing the second order network dependencies. As might
be expected, the standard regression model performs substantially worse
than the others.
Finally, it is interesting to examine the estimates of the $\beta$'s for
the standard regression model against those of the LSR model, which
accounts for the temporal and network dependence inherent in the data.
The results are shown graphically in Figure \ref{BetaLsnReg}. The
figure illustrates two main points: (1) Even though the model for the
expected value, unconditional on the random effects, is the same
($x'_{i,j,t} \beta_t$), there is a definite difference in the estimated
values of the coefficients; (2) The 95\% credible intervals for
standard regression are generally shorter than those for the LSR model.
However, the length of the intervals of the $\beta$'s for the polity of
the exporter, cooperation in conflict and polity interaction are
actually shorter for the LSR model compared to those of the standard
regression model. These results illustrate that accounting for
dependency in data typically increases the nominal precision of the
estimated coefficients, but this is not always the case,
and depends on the distribution of the covariates
themselves.\looseness=1

%s6 ###
\section{Militarized interstate disputes}\label{sec6}

Jones, Bremer and Singer (\citeyear{JonBreSin96}) defined the term \textit{militarized interstate
dispute} (MID) as an event ``in which the threat, display or use of
military force short of war by one member state is explicitly directed
toward the government, official representatives, official forces,
property, or territory of another state.'' In this analysis, we will
investigate the patterns of MIDs in the Middle East and United States
from 1991 to 2000.\footnote{A list of countries used in this analysis
(including their three-letter ISO code) can be found in the \hyperref[appm]{Appendix}.
Additionally, the data and some of the R code used to fit the model are
available as supplementary material [\citet{WesHofLSNSup}].} For this
data analysis, $y_{i,j,t}$ is the binary indicator of a MID initiated
by country $i$ with target $j$ in year $t$. We are interested in
relating the response to the following covariates: (1) the ordinal
level of alliance between $i$ and $j$, ranging from 0 (no alliance) to
3 (will defend each other militarily), (2) the real value of the log
trade from $i$ to $j$, (3) the real value of the log trade from $j$ to
$i$, (4) the number of inter-governmental associations of which both
nations are members, and (5) the log distance between the two nations.
Note that all of the covariates, except distance, are potentially
changing over time.

As discussed in Section~\ref{sec4.2}, for the probit mixed effects model the
variance of $g_{i,j,t}$ is set to one, leading to additional
restrictions on the priors for $\phi_g$, $\phi_{gg}$ and $\rho_{gg}$.
Specifically, we considered the following set of diffuse priors:
\begin{eqnarray*}
\beta& \sim&
\operatorname{mvn}(0, 100 \times\mathrm{I}), \\
(\phi_s, \phi_{sr}, \phi_{rs}, \phi_r)' & \sim& \operatorname{mvn}(0, 100
\times\mathrm{I}) \mathbb{I}(\Phi_{sr} \in\mathcal{S}), \\
(\phi_g, \phi_{gg})' & \sim& \operatorname{mvn}(0, 100 \times\mathrm{I})
\mathbb{I}(\Phi_{gg} \in\mathcal{S}) \mathbb{I}(\Gamma_{gg} \mbox{
is positive definite}),
\\
\Gamma_{sr} & \sim& \operatorname{inverse\mbox{-}Wishart}(4, \mathrm{I}),\\
\rho_{gg} &\sim& \operatorname{normal}(0, 100) \mathbb{I}(-1 \leq \rho_{gg}
\leq1) \mathbb{I} (\Gamma_{gg} \mbox{ is positive definite}).
\end{eqnarray*}
The posterior distribution for these parameters was approximated
with\break
a~Markov chain Monte Carlo algorithm consisting of 7 million scans. The
first two million of these scans were dropped to allow for convergence
to the stationary distribution. Parameter values were saved every
1,000th scan, resulting in 5,000 samples for each
parameter with which to approximate the posterior distribution.

%f5 ###
\begin{figure}

\includegraphics{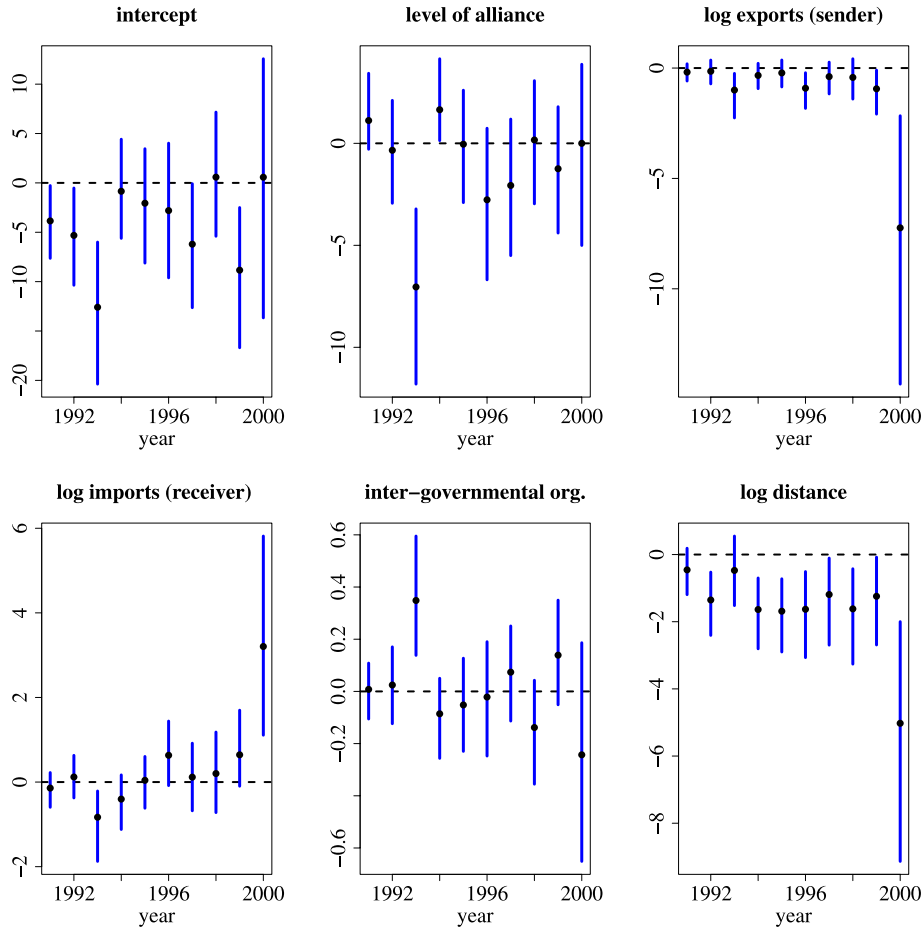}

\caption{95\% posterior credible intervals for the covariate coefficients.}\vspace*{-4pt}
\label{MIDBeta}
\end{figure}
%f6 ###
\begin{figure}

\includegraphics{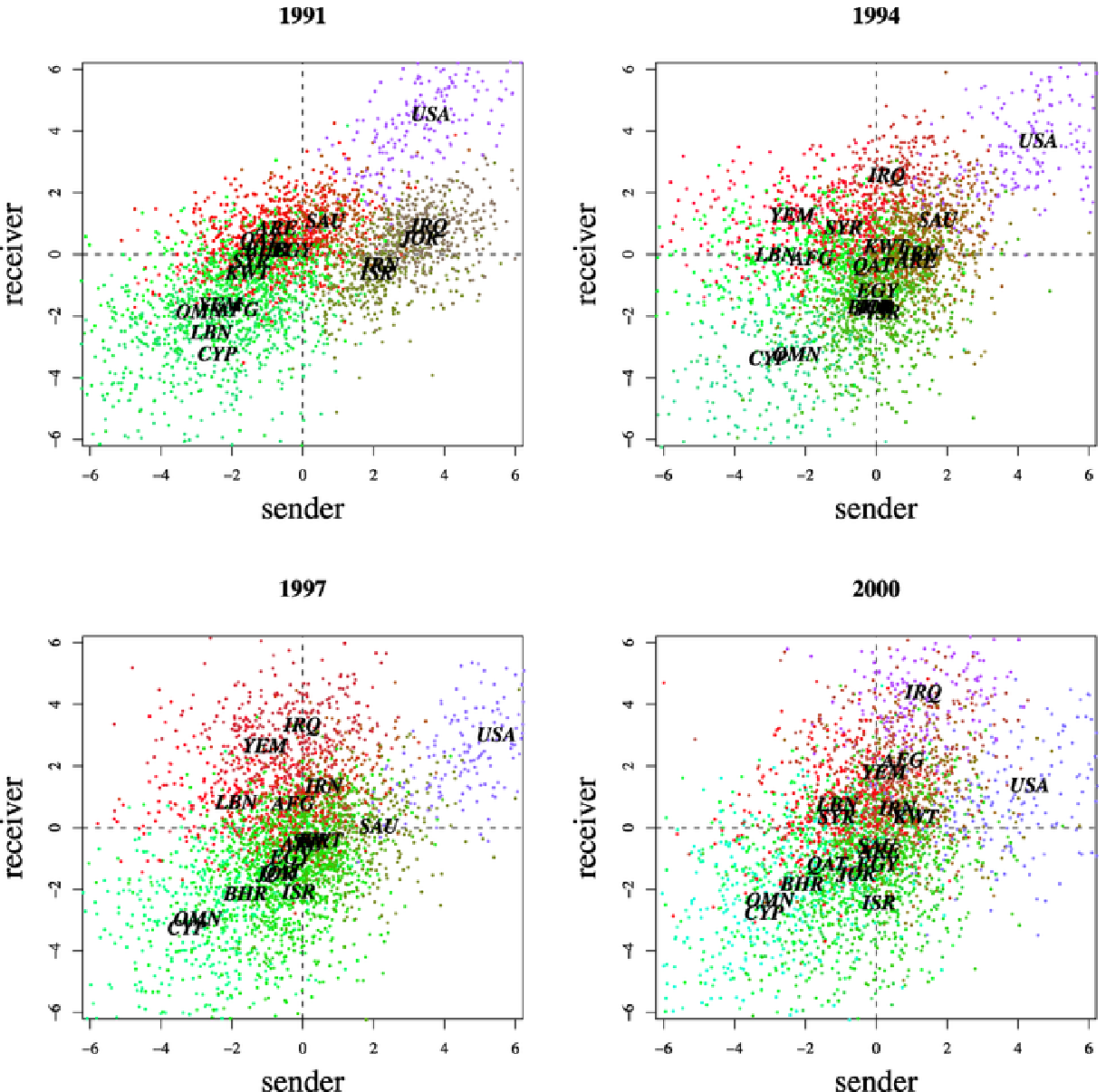}

\caption{Sender--receiver effects for the model with covariates.}
\label{MIDSR}
\end{figure}

%%%%%%%%%%%%%%%
%s6.1 ###
\subsection{Results}\label{sec6.1}
Figure \ref{MIDBeta} presents the 95\% credible intervals for the
coefficients of the covariates. We focus attention on the intervals not
containing zero (with high credibility), suggesting an effect on MIDs.
Overall, the pattern of the intervals for the level of alliance between
a pair of nations appears mixed. As might be expected, for four of the
years (1993, 1996, 1997, 1999) the medians are below zero, suggesting a
negative impact on MIDs for higher levels of alliance---in 1999, the
empirical probability that the coefficient is below zero is 79\%.
However, in 1991 and 1994, it appears that the higher the level of
alliance between two nations, the more likely that they would have a
MID. A possible reason for this paradox is that Oman has the only level
3 alliances in the data and in 1991 it had disputes with both Iraq and
Jordan, and in 1994 it had a dispute with Iraq.\vadjust{\goodbreak} The effect of the
number of inter-governmental organizations to which a pair of nations
belong also appears to be minimal over the period, except for the year
1993. The effects of the log of exports from the initiator to the
target and the log of imports from the target to the initiator are very
interesting. Over the period, there appears to be a slight trend for
the coefficients of both of the covariates (with extremely large
variability in the final year).\footnote{We also fit the model without
the year 2000 and found the precision of the $\beta$'s for 1991--1999 to
be similar to those in Figure \ref{MIDBeta}.} These trends suggest that
the more a potential initiator of a dispute exports to a particular
nation, the less likely it is for a dispute to occur. This is in
contrast to importing from a particular country. Finally, distance
appears to be a deterrent to conflict; the farther a pair of nations
are from each other, the smaller the chance of a militarized dispute
between the pair.\vadjust{\goodbreak}

\begin{figure}

\includegraphics{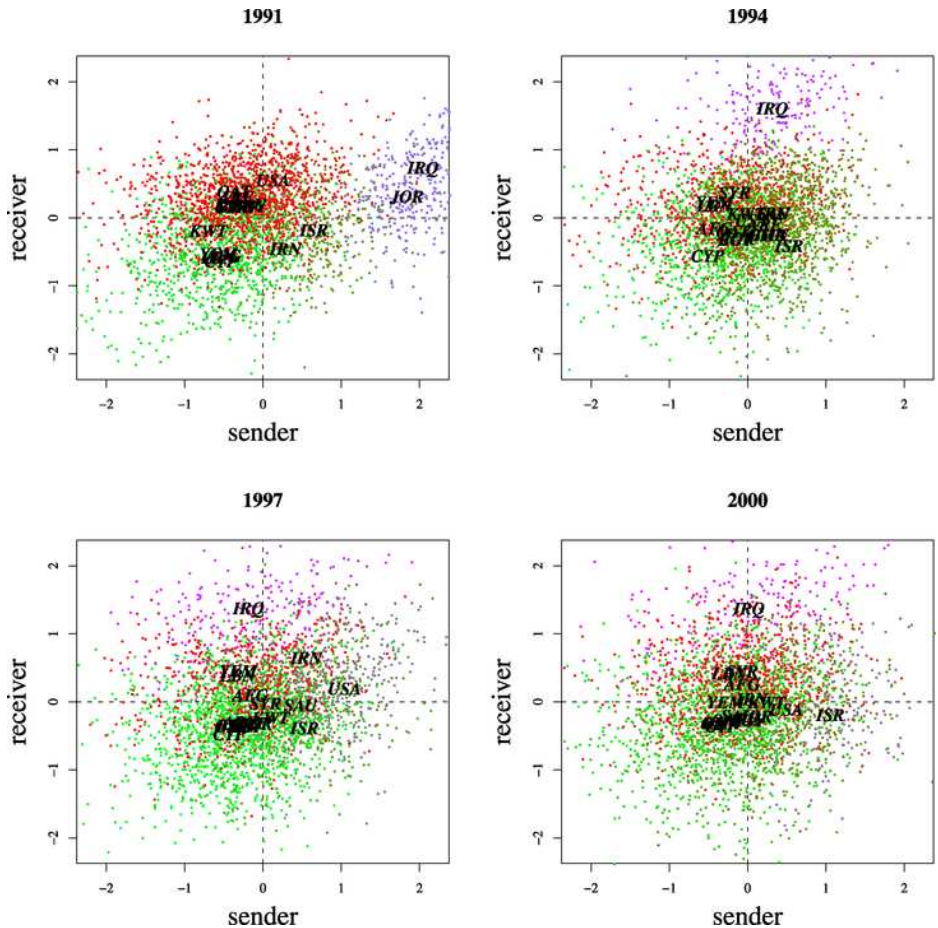}

\caption{Sender--receiver effects for the mean model.}
\label{MIDMuSR}%
\end{figure}

Figure \ref{MIDSR} presents 200 random samples from the bivariate
posterior distribution of the sender and receiver effects for each
country. These effects represent deviations of the country-specific
rates of initiating and receiving MIDs from what would be predicted by
a probit regression model alone. From the figure, we see that there are
some nations for which the distributions do not overlap, suggesting
differences between the nations with respect to their sending and
receiving effects. There appears to be a positive correlation between
the sending and receiving of militarized disputes---the median
correlation turns out to be 0.563 (Table \ref{BetaGammaMID}). As the
United States (USA) is near the far right corner for all the plots in
the figure, it is the most active in the network over the period. This
suggests that the United States has far more disputes than would be
expected, given just its covariate information. In particular, since
distance is generally a significant deterrent to disputes, the United
States has far more disputes than would be expected, based on its
distance from the Middle East. Note that in 1991, Iraq (IRQ) and Jordan
(JOR) are also high initiators of disputes. However, Oman (OMN),
Lebanon (LBN) and Cyprus (CYP) neither initiate nor are the target of
many disputes over the period. In contrast, the results can be compared
to those in Figure \ref{MIDMuSR}, where the analysis was conducted
without the covariates; that is, only a mean was fit at each point in
time ($y_{i,j,t} = \mu_t + s_{i,t} + r_{j,t} + g_{i,j,t}$). Now the
United States is no longer in the upper right-hand corner of the plots.
However, Cyprus is still in the lower left-hand corner of all the
plots. In this case, accounting for influential covariate information
induces greater variability in the sender--receiver random effects.

%t6 ###
\begin{table}%[!htb]
\caption{$\Sigma(0)_{sr}$ and $\Sigma(0)_{gg}$ parameter estimates for
the MIDs model}
\label{BetaGammaMID}
\vspace*{-3pt}
\begin{tabular}{@{}lcccc@{}}
\hline
\textbf{Parameter} &
\textbf{Markov chain} &
\multicolumn{1}{c}{\textbf{2.5\%}} &
\multicolumn{1}{c}{\textbf{Median}} &
\multicolumn{1}{c@{}}{\textbf{97.5\%}}
\\
\hline
$\sigma^2_s$ &
\includegraphics{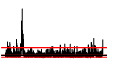}  &
$\textcolor{red}{1.632}$ &
$\textcolor{black}{5.412}$ &
$\textcolor{red}{21.088}$
\\
$\sigma_{{sor}}$ &
\includegraphics{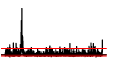} &
$\textcolor{red}{0.314}$ &
$\textcolor{black}{2.831}$ &
$\textcolor{red}{14.373}$
\\
$\rho_{sr}$ &
\includegraphics{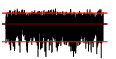} &
$\textcolor{red}{0.100}$ &
$\textcolor{black}{0.563}$ &
$\textcolor{red}{\phantom{2}0.846}$
\\
$\sigma^2_r$ &
\includegraphics{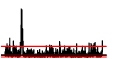} &
$\textcolor{red}{1.765}$ &
$\textcolor{black}{5.491}$ &
$\textcolor{red}{23.056}$
\\
$\sigma^2_g \equiv1$ &
\includegraphics{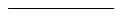} &
$\textcolor{red}{1\phantom{.632}}$ &
$\textcolor{black}{1\phantom{.683}}$ &
$\textcolor{red}{1\phantom{.73}}$
\\
$\rho_{gg}$ &
\includegraphics{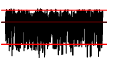} &
$\textcolor{red}{0.187}$ &
$\textcolor{black}{0.683}$ &
$\textcolor{red}{\phantom{2}0.955}$
\\
\hline
\end{tabular}
\vspace*{-5pt}
\end{table}
%

%f7 ###

Tables \ref{BetaGammaMID} and \ref{BetaARMID} describe the variability
of the sender and receiver effects and the temporal variation. The
median of $\rho_{gg}$ is 0.683, and while the 95\% credible interval is
quite spread out, its range is completely above zero, suggesting a
certain amount of positive reciprocity in the network at a given point
in time. However, since the median of the posterior distribution
of~$\phi_{gg}$ is 0.189 (and this interval partially contains zero), we
see that positive reciprocity in a given year may not be readily
explained by the level of reciprocity in the previous year. Since the
median of the posterior distributions for $\phi_{s}$ and $\phi_{sr}$
are 0.761 and 0.245, respectively, we see that the initiation of
disputes by a particular nation depends to a large degree on whether
they initiated disputes in the previous year and, to a lesser extent,
on whether they were a target in the previous year. Finally, the median
of the posterior distributions for $\phi_{r}$ and $\phi_{rs}$ are 0.909
and $-$0.029, respectively. This suggests that whether a nation is a
target this year depends heavily on whether they were a target in the
previous year, but depends very little on whether they initiated
disputes in the previous year.

%%%%%%%%%%%%%%%%%%%%%%%%%%%%%%%%%%
%s7 ###
\section{Discussion}\label{sec7}
This paper has developed a framework that incorporates temporal
dependence within the domain of social relations regression models. We
showed that our particular mixed effects model can account for both
second order network dependence and temporal dependence. By placing the
temporal dependence\vadjust{\eject} on the random effects representing the network
dependence, the network is allowed to evolve over time. Additionally, a
generalized linear modeling framework was developed and a general
Bayesian estimation approach was outlined. Specific examples for
Gaussian and binary responses were illustrated and applied to the study
of international trade and militarized interstate disputes,
respectively. The incorporation of temporal dependence allowed for
insight into the network of international trade by noting that after
accounting for covariate information, the level of exports in a given
year is highly dependent on the level from the previous year, but not
dependent on the level of imports. Conversely, the level of imports in
a~particular year is fairly dependent on imports from the previous year
and only somewhat dependent on exports the previous year. Additionally,
only a slight degree of reciprocity can be explained by the level of
reciprocity in the previous year.

%t7 ###
\begin{table}%[!htb]
\caption{$\Phi_{sr}$ and $\Phi_{gg}$ parameter estimates for the MIDs model}
\label{BetaARMID}
\begin{tabular}{@{}lcd{2.3}d{2.3}c@{}}
\hline
\textbf{Parameter}&\textbf{Markov chain}&\multicolumn{1}{c}{\textbf{2.5\%}}&\multicolumn{1}{c}{\textbf{Median}}&\textbf{97.5\%}\\
\hline
$\phi_s$ &
\includegraphics{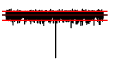}
 & $\textcolor{red}{0.560}$ & $\textcolor{black}{0.761}$ & $\textcolor{red}{0.902}$ \\
$\phi_{sr}$ &
\includegraphics{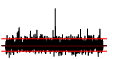}
 & $\textcolor{red}{0.067}$ & $\textcolor{black}{0.245}$ & $\textcolor{red}{0.468}$ \\
$\phi_{rs}$ &
\includegraphics{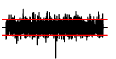}
 & $\textcolor{red}{-0.207}$ & $\textcolor{black}{-0.029}$ & $\textcolor{red}{0.155}$ \\
$\phi_r$ &
\includegraphics{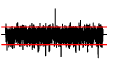}
 & $\textcolor{red}{0.679}$ & $\textcolor{black}{0.909}$ & $\textcolor{red}{1.081}$ \\
$\phi_g$ &
\includegraphics{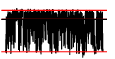}
 & $\textcolor{red}{0.003}$ & $\textcolor{black}{0.741}$ & $\textcolor{red}{0.947}$ \\
$\phi_{gg}$ &
\includegraphics{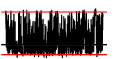}
 & $\textcolor{red}{-0.034}$ & $\textcolor{black}{0.189}$ & $\textcolor{red}{0.920}$ \\
\hline
\end{tabular}
\end{table}

The authors plan to extend the research by (1) considering other
approaches for modeling the temporal dependence, both stationary and
nonstationary, and (2) allowing for third-order dependencies, such as
those outlined in \citet{Hof02} and \citet{Hofbi}, to be dependent over time.

\begin{appendix}
%s8 ###
\section{\texorpdfstring{MCMC algorithm for the Gaussian case}{Appendix A: MCMC algorithm for the Gaussian case}}\label{appm}
Parameter estimation is conducted through the construction of a Markov
chain in the parameters $\Theta$. The following MCMC algorithm presents
one possible construction:
\begin{enumerate}[6.]
\item Sample $\beta$ from its $\operatorname{mvn}(M, V)$ full conditional
distribution, where
\begin{eqnarray*}
V &=&  \Biggl(\sum_{i=1}^{A-1} \sum_{j=i+1}^A x_{[i,j]}'\Sigma
_{gg}^{-1}x_{[i,j]} + V_{\beta}^{-1} \Biggr)^{-1},\\
M &=& V \Biggl(\sum_{i=1}^{A-1} \sum_{j=i+1}^A x_{[i,j]}'\Sigma
_{gg}^{-1}B_{[i,j]} + V_{\beta}^{-1}M_{\beta} \Biggr), \\
B_{[i,j]} &=&
y_{[i,j]} - sr_i - rs_j.
\end{eqnarray*}
\item Sample each $sr_i$, $i \in1, \ldots, A$, from its
$\operatorname{mvn}(M, V)$
full conditional distribution, where
\begin{eqnarray*}
V &=&  \bigl((A-1)\Sigma_{gg}^{-1} + \Sigma_{sr}^{-1} \bigr)^{-1}, \\
M &=& V  \Biggl(\Sigma_{gg}^{-1}\sum_{j\not=i =1}^A B_{[i,j]} \Biggr), \\
B_{[i,j]} &=& y_{[i,j]} - x_{[i,j]}\beta- rs_j.
\end{eqnarray*}
\item Sample $(\phi_s^*, \phi_{sr}^*, \phi_{rs}^*, \phi_r^*)'$ from
a $\operatorname{mvn}(M, V)$ distribution, where
\begin{eqnarray*}
V &=&  \Biggl(\sum_{i=1}^A \sum_{t=1}^T Z_{i,t-1}' \Gamma_{sr}^{-1}
Z_{i,t-1} + V_{\Phi_{sr}}^{-1}  \Biggr)^{-1}, \\
M &=& V  \Biggl(\sum_{i=1}^A \sum_{t=1}^T Z_{i,t-1}' \Gamma_{sr}^{-1}
sr_{i,t} + V_{\Phi_{sr}}^{-1}M_{\Phi_{sr}} \Biggr), \\
Z_{i,t-1} &=&
\pmatrix{\displaystyle
sr_{i,t-1}' & 0 \cr\displaystyle
0 & sr_{i,t-1}'
}
.
\end{eqnarray*}
\begin{enumerate}[(b)]
\item[(a)] Calculate $\Sigma_{sr}^*$ from $\Phi_{sr}^*$ and $\Gamma_{sr}$
and compute the Metropolis--Hastings ratio:
\begin{eqnarray*}
r &=& \frac{ \prod_{i=1}^A \operatorname{dmvn} (sr_i | 0; \Sigma
_{sr}^*(\Phi_{sr}^*, \Gamma_{sr}) )}{\prod_{i=1}^A \operatorname
{dmvn} (sr_i | 0; \Sigma_{sr}(\Phi_{sr}, \Gamma_{sr}) )}
\\
& & {}\times\frac{ \operatorname{dmvn} ( \Phi_{sr}^* | M_{\Phi_{sr}}; V_{\Phi
_{sr}} )}{\operatorname{dmvn} ( \Phi_{sr} | M_{\Phi_{sr}}; V_{\Phi
_{sr}} )}
\times\frac{ \operatorname{dmvn} ( \Phi_{sr} | M; V  )}{\operatorname
{dmvn} ( \Phi_{sr}^* | M; V )}.
\end{eqnarray*}
\item[(b)] Accept $\Phi_{sr}^*$ with probability $r \wedge1.$
\end{enumerate}
\item Sample $(\phi_g^*, \phi_{gg}^*)'$ from
a $\operatorname{mvn}(M, V)$ distribution, where
\begin{eqnarray*}
V &=&
\Biggl(\sum_{i=1}^{A-1} \sum_{j=i+1}^A \sum_{t=1}^T Z_{[i,j],t-1}'
\Gamma_{gg}^{-1} Z_{[i,j],t-1} + V_{\Phi_{gg}}^{-1}  \Biggr)^{-1}, \\
M &=& V \Biggl (\sum_{i=1}^{A-1} \sum_{j=i+1}^A \sum_{t=1}^T
Z_{[i,j],t-1}' \Gamma_{gg}^{-1} g_{[i,j],t} + V_{\Phi_{gg}}^{-1}M_{\Phi
_{gg}} \Biggr) , \\
g_{[i,j],t} &=& y_{[i,j],t} - \eta_{[i,j],t} - sr_{i,t} - rs_{j,t},\\
Z_{[i,j],t-1} &=&  \pmatrix{\displaystyle
g_{i,j,t-1} & g_{j,i,t-1} \cr\displaystyle
g_{j,i,t-1} & g_{i,j,t-1}
}
.
\end{eqnarray*}
\begin{enumerate}[(b)]
\item[(a)] Calculate $\Sigma_{gg}^*$ from $\Phi_{gg}^*$ and $\Gamma_{gg}$
and compute the Metropolis--Hastings ratio:
\begin{eqnarray*}
r &=& \frac{ \prod_{i=1}^{A-1} \prod_{j=i+1}^A \operatorname{dmvn}(g_{[i,j]}
| 0; \Sigma_{gg}^*(\Phi_{gg}^*, \Gamma_{gg}))}{\prod_{i=1}^{A-1} \prod
_{j=i+1}^A \operatorname{dmvn}(g_{[i,j]} | 0; \Sigma_{gg}(\Phi_{gg}, \Gamma
_{gg}))} \\
& &{} \times\frac{ \operatorname{dmvn} ( \Phi_{gg}^* | M_{\Phi_{gg}}; V_{\Phi
_{gg}} )}{\operatorname{dmvn} ( \Phi_{gg} | M_{\Phi_{gg}}; V_{\Phi
_{gg}} )}
\times\frac{ \operatorname{dmvn} ( \Phi_{gg} | M; V  )}{\operatorname
{dmvn} ( \Phi_{gg}^* | M; V )}.
\end{eqnarray*}
\item[(b)] Accept $\Phi_{gg}^*$ with probability $r \wedge1.$
\end{enumerate}
\item Sample $\Gamma_{sr}^*$ from an inverse\mbox{-}Wishart$(AT + v_{sr}, \{
SS_{sr} + S_{sr}\}^{-1})$ distribution, where
\begin{eqnarray*}
SS_{sr} &=& \sum_{i=1}^A \sum_{t=1}^T  (sr_{i,t} - \Phi
_{sr}sr_{i,t-1} ) (sr_{i,t} - \Phi_{sr}sr_{i,t-1} )'.
\end{eqnarray*}
\begin{enumerate}[(b)]
\item[(a)] Calculate $\Sigma_{sr}^*$ from $\Phi_{sr}$ and $\Gamma_{sr}^*$
and compute the Metropolis--Hastings ratio:
\begin{eqnarray*}
r &=& \frac{ \prod_{i=1}^A \operatorname{dmvn} (sr_i | 0; \Sigma
_{sr}^*(\Phi_{sr}, \Gamma_{sr}^*) )}{\prod_{i=1}^A \operatorname
{dmvn} (sr_i | 0; \Sigma_{sr}(\Phi_{sr}, \Gamma_{sr}) )}\\
&&{}\times \frac{ \operatorname{inverse\mbox{-}Wishart} ( \Gamma_{sr}^* | v_{sr};
S_{sr}^{-1} )}{\operatorname{inverse\mbox{-}Wishart} ( \Gamma_{sr} |
v_{sr}; S_{sr}^{-1} )}\\
&&{}\times \frac{ \operatorname{inverse\mbox{-}Wishart} ( \Gamma_{sr} | AT +
v_{sr}; \{SS_{sr} + S_{sr}\}^{-1} )}{\operatorname{inverse\mbox{-}Wishart}
( \Gamma_{sr}^* | AT + v_{sr}; \{SS_{sr} + S_{sr}\}^{-1} )}.
\end{eqnarray*}
\item[(b)] Accept $\Gamma_{sr}^*$ with probability $r \wedge1.$
\end{enumerate}
\item Sample a proposal for $\Gamma_{gg}^*$ as follows:
\begin{eqnarray*}
 [\sigma_{a}^{2*}| \cdot ]
 &\sim& \operatorname{inverse\mbox{-}gamma}\Biggl(
\alpha_a^{\dagger}= N/2 + \alpha_a, \delta_a^{\dagger} = \Biggl(\sum_{m=1}^N
a_m^2\Biggr)\Big/2 + \delta_a\Biggr), \\
 {[}\sigma_{b}^{2*}| \cdot ]
 &\sim& \operatorname{inverse\mbox{-}gamma}
\Biggl(\alpha_b^{\dagger}=N/2 + \alpha_b, \delta_b^{\dagger} =\Biggl(\sum_{m=1}^N
b_m^2\Biggr)\Big/2 +\delta_b\Biggr),
\end{eqnarray*}
where $N = A(A-1)(T-1)$,
and set $\gamma^{2*}_{g}= ( \sigma^{2*}_a+ \sigma^{2*}_b )/4$ and
$\lambda_{gg}^* = ( \sigma^{2*}_a- \sigma^{2*}_b )/(\sigma^{2*}_a+\sigma
^{2*}_b)$.
\begin{enumerate}[(b)]
\item[(a)] Calculate $\Sigma_{gg}^*$ from $\Phi_{gg}$ and $\Gamma_{gg}^*$
and compute the Metropolis--Hastings ratio:
\begin{eqnarray*}
r &=& \frac{ \prod_{i=1}^{A-1} \prod_{j=i+1}^A \operatorname{dmvn}(g_{[i,j]}
| 0; \Sigma_{gg}^*(\Phi_{gg}, \Gamma_{gg}^*))}{\prod_{i=1}^{A-1} \prod
_{j=i+1}^A \operatorname{dmvn}(g_{[i,j]} | 0; \Sigma_{gg}(\Phi_{gg}, \Gamma
_{gg}))}\\
&&{}\times \frac{\operatorname{inverse\mbox{-}gamma}(\sigma^{2*}_a | \alpha_a;\delta
_a)}{ \operatorname{inverse\mbox{-}gamma}(\sigma^{2}_a| \alpha_a; \delta_a)} \\
&&{}\times \frac{\operatorname{inverse\mbox{-}gamma}(\sigma^{2*}_b | \alpha_b; \delta
_b)}{ \operatorname{inverse\mbox{-}gamma}(\sigma^{2}_b| \alpha_b; \delta_b)} \\
&& {}\times \frac{\operatorname{inverse\mbox{-}gamma}(\sigma^2_a | \alpha_a^{\dagger
}; \delta_a^{\dagger})}{ \operatorname{inverse\mbox{-}gamma}(\sigma^{2*}_a| \alpha
_a^{\dagger}; \delta_a^{\dagger})} \\
&&{} \times \frac{\operatorname{inverse\mbox{-}gamma}(\sigma^2_b | \alpha_b^{\dagger
}; \delta_b^{\dagger})}{ \operatorname{inverse\mbox{-}gamma}(\sigma^{2*}_b| \alpha
_b^{\dagger}; \delta_b^{\dagger})}.
\end{eqnarray*}
\item[(b)] Accept $\Gamma_{gg}^*$ with probability $r \wedge1.$
\end{enumerate}
\item Sample the missing data $y_{i,j,t}$ from its $\operatorname{mvn}(M_t,V_t)$ full conditional distribution,
where
\begin{enumerate}[(b)]
\item[(a)]$t=1$:
\begin{eqnarray*}
\mu_{[i,j],1}&=& \eta_{[i,j],1} + sr_{i,1} + rs_{j,1},\\
C_{[i,j],2} &=& g_{[i,j],2} + \Phi_{gg} \mu_{[i,j],1},\\
V_1 &=& \bigl(\Phi_{gg}' \Gamma_{gg}^{-1} \Phi_{gg} + \Sigma
_{gg}(0)^{-1}\bigr)^{-1},\\
M_1 &=& V_1\bigl(\Phi_{gg}' \Gamma_{gg}^{-1} C_{[i,j],2} + \Sigma
_{gg}(0)^{-1} \mu_{[i,j],1}\bigr).
\end{eqnarray*}
\item[(b)]$1< t < T$:
\begin{eqnarray*}
\mu_{[i,j],t} &=& \eta_{[i,j],t} + sr_{i,t} + rs_{j,t},\\
C_{[i,j],t+1} &=& g_{[i,j],t+1} + \Phi_{gg} \mu_{[i,j],t},\\
D_{[i,j],t-1} &=& \mu_{[i,j],t} + \Phi_{gg} g_{[i,j],t-1},\\
V_t &=& (\Phi_{gg}' \Gamma_{gg}^{-1} \Phi_{gg} + \Gamma
_{gg}^{-1})^{-1},\\
M_t &=& V_t\bigl(\Phi_{gg}' \Gamma_{gg}^{-1} C_{[i,j],t+1} + \Gamma
_{gg}^{-1} D_{[i,j],t-1}\bigr).
\end{eqnarray*}
\item[(c)]$t=T$:
\begin{eqnarray*}
\mu_{[i,j],T} &=& \eta_{[i,j],T} + sr_{i,T} + rs_{j,T},\\
D_{[i,j],T-1} &=& \mu_{[i,j],T} + \Phi_{gg} g_{[i,j],T-1}, \\
V_T &=& \Gamma_{gg},\\
M_T &=& V_T\bigl(\Gamma_{gg}^{-1} D_{[i,j],T-1}\bigr).
\end{eqnarray*}
\end{enumerate}
\end{enumerate}

%s9 ###
\section{\texorpdfstring{MCMC algorithm for the probit model}{Appendix B: MCMC algorithm for the probit model}}
In order to augment the previous algorithm for the probit LSR model,
the Gibbs sampling procedure for each $(i,j)$ and $(j,i)$ pair at the
following times $t=1,\ldots,T$ proceeds by sampling the conditional
distribution for each~$\theta_{i,j,t}$, based on a truncated normal
distribution; the truncation is to the left of zero if $y_{i,j,t}=0$
and to the right of zero if $y_{i,j,t}=1$:
\begin{eqnarray*}
 [\theta_{[i,j],t}| \cdot ]
 &\sim& \operatorname{mvn}(M_t, V_t),\\
 {[}\theta_{i,j,t} | \theta_{j,i,t}, \cdot ]
 &\sim&
\cases{\displaystyle
\operatorname{normal}(M_t^*, V_t^*)\mathbb{I}( \theta_{i,j,t} < 0)\mathbb
{I}({y_{i,j,t}=0}),\cr\displaystyle
\operatorname{normal}(M_t^*, V_t^*)\mathbb{I}( \theta_{i,j,t} > 0)\mathbb
{I}({y_{i,j,t}=1}),
}
 \\
 {[}\theta_{j,i,t} | \theta_{i,j,t}, \cdot ]
  &\sim&
\cases{\displaystyle
\operatorname{normal}(M_t^*, V_t^*)\mathbb{I}( \theta_{j,i,t} < 0)\mathbb
{I}({y_{j,i,t}=0}),\cr\displaystyle
\operatorname{normal}(M_t^*, V_t^*)\mathbb{I}( \theta_{j,i,t} > 0)\mathbb
{I}({y_{j,i,t}=1}).
}
\end{eqnarray*}
The means and variances of $ [\theta_{[i,j],t}| \cdot ]$ for
$t =1, 1< t < T, t=T$ have the same expressions as
those for $y_{i,j,t}$ in Step 7 of
Appendix~\ref{appm}.
%
%%
%%
%C_{[i,j],2} &=& g_{[i,j],2} + \Phi_{gg} \mu_{[i,j],1},\\
%V_1 &=& \bigl(\Phi_{gg}' \Gamma_{gg}^{-1} \Phi_{gg} + \Sigma
%_{gg}(0)^{-1}\bigr)^{-1},\\
%M_1 &=& V_1\bigl(\Phi_{gg}' \Gamma_{gg}^{-1} C_{[i,j],2} + \Sigma
%_{gg}(0)^{-1} \mu_{[i,j],1}\bigr).
%%
%%
%C_{[i,j],t+1} &=& g_{[i,j],t+1} + \Phi_{gg} \mu_{[i,j],t},\\
%D_{[i,j],t-1} &=& \mu_{[i,j],t} + \Phi_{gg} g_{[i,j],t-1},\\
%V_t &=& (\Phi_{gg}' \Gamma_{gg}^{-1} \Phi_{gg} + \Gamma
%_{gg}^{-1})^{-1},\\
%M_t &=& V_t\bigl(\Phi_{gg}' \Gamma_{gg}^{-1} C_{[i,j],t+1} + \Gamma
%_{gg}^{-1} D_{[i,j],t-1}\bigr).
%%
%%
%D_{[i,j],T-1} &=& \mu_{[i,j],T} + \Phi_{gg} g_{[i,j],T-1}, \\
%V_T &=& \Gamma_{gg},\\
%M_T &=& V_T\bigl(\Gamma_{gg}^{-1} D_{[i,j],t-1}\bigr).
%%
%
For $\rho_{gg}$ we simply suggest using a Metropolis--Hastings update
using an uniform proposal distribution around the current value.
The range of this distribution is the only tuning parameter in the
Markov chain Monte Carlo algorithm.

%s10 ###
\section{\texorpdfstring{Set of nations in the trade application}{Appendix C: Set of nations in the trade application}}
Algeria (DZA), Argentina (ARG), Australia (AUS),
Austria (AUT), Barbados (BRB), Belgium (BEL),
Bolivia (BOL), Brazil (BRA), Canada (CAN),
Chile (CHL), Colombia (COL), Costa Rica (CRI),
Cyprus (CYP), Denmark (DNK), Ecuador (ECU),
Egypt (EGY), El Salvador (SLV), Finland (FIN),
France (FRA), Germany (DEU), Greece (GRC),
Guatemala (GTM), Honduras (HND), Iceland (ISL),
India (IND), Indonesia (IDN), Ireland (IRL),
Israel (ISR), Italy (ITA), Jamaica (JAM),
Japan (JPN), Malaysia (MYS),
Mauritius (MUS), Mexico (MEX), Morocco (MAR),
Nepal (NPL), Netherlands (NLD), New Zealand (NZL),
Norway (NOR), Oman (OMN), Panama (PAN),
Paraguay (PRY), Peru (PER), Philippines (PHL),
Portugal (PRT), Republic of Korea (KOR), Singapore (SGP), Spain (ESP),
Sweden (SWE), Switzerland (CHE), Thailand (THA),
Trinidad and Tobago (TTO), Tunisia (TUN), Turkey (TUR),
United Kingdom (GBR), United States (USA), Uruguay (URY),
Venezuela (VEN).

%s11 ###
\section{\texorpdfstring{Set of nations in the MIDs application}{Appendix D: Set of nations in the MIDs application}}
Afghanistan (AFG), Bahrain (BHR), Cyprus (CYP), Egypt (EGY), Iran
(IRN), Iraq (IRQ), Israel (ISR), Jordan (JOR), Kuwait (KWT), Lebanon
(LBN), Oman (OMN), Qatar (QAT), Saudi Arabia (SAU), Syria (SYR), United
Arab Emirates (ARE), United States (USA), and Yemen (YEM).
\end{appendix}

%%%%%%%%%%%%%%%%%%%%%%%%%%%%%%%%
\section*{Acknowledgments}
The authors would like to thank the Associate Editor and referees for
their advice on this manuscript.
Additionally, we are appreciative to both Michael D. Ward and Xun Cao
for collecting and supplying the
data used in this paper. Finally, the corresponding author would also
like to thank
Grace S. Chiu, Patrick J. Heagerty, Kevin M. Quinn and Michael D. Ward
for enlightening
discussions on this topic.

%%%%%%%%%%%%%%%%%%%%%%%%%%%%%%%%
%
\begin{supplement}[id=suppA]
\stitle{Data and R Code for the Examples}
\slink[doi]{10.1214/10-AOAS403SUPP} %[doi,text={...}] - jei reikia
%suskaldyti doi
\slink[url]{http://lib.stat.cmu.edu/aoas/403/supplement.zip}
\sdatatype{.zip}
\sdescription{A zip file associated with the paper contains the data
and some of the R code used in the examples.}
\end{supplement}

%suskaldyti doi

\printaddresses

\end{document}